\documentclass[12pt]{iopart}
\pdfoutput=1
\usepackage{graphicx} 
\usepackage{xcolor}
\usepackage{lineno}
\usepackage{cite}
\usepackage{xspace}
\usepackage{amssymb}
\usepackage{url}
\usepackage[T1]{fontenc}

\begin{document}

\title[]{Identification of low energy nuclear recoils in a gas TPC with optical readout}

\newcommand {\ie}{\mbox{i.e.}\xspace}     
\newcommand {\eg}{\mbox{e.g.}\xspace}     

\newcommand{\fe}{\ensuremath{^{55}\textrm{Fe}}\xspace}
\newcommand{\abs}[1]{\ensuremath{\vert #1 \vert}}
\newcommand{\ambe}{\ensuremath{\textrm{Am} \textrm{Be}}\xspace}
\newcommand{\isclu}{\ensuremath{I_{SC}}\xspace}
\newcommand{\tsigmag}{\ensuremath{\sigma^T_{Gauss}}\xspace}
\newcommand{\dedl}{\ensuremath{\frac{dE}{dl_p}}\xspace}

\newcommand{\lemon}{{\textsc{Lemon}}\xspace}
\newcommand{\cygno}{{\textsc{Cygno}}\xspace}
\newcommand{\idbscan}{{\textsc{Idbscan}}\xspace}
\newcommand{\dbscan}{{\textsc{dbscan}}\xspace}
\newcommand{\gac}{{\textsc{Gac}}\xspace}
\newcommand{\nnc}{{\textsc{Nnc}}\xspace}
\newcommand{\GEANTfour} {{\textsc{Geant4}}\xspace}
\newcommand{\SRIM} {{\textsc{Srim}}\xspace}
\newcommand{\garfield} {{\textsc{Garfield}}\xspace}
\newcommand{\PYTHONthree} {{\textsc{Python3}}\xspace}
\newcommand{\ROOT} {{\textsc{Root}}\xspace}

\newcommand{\unit}[1]{\ensuremath{\textrm{\,#1}}\xspace}
\newcommand{\keV}{\ensuremath{\,\textrm{ke\hspace{-.08em}V}}\xspace}
\newcommand{\MeV}{\ensuremath{\,\textrm{Me\hspace{-.08em}V}}\xspace}



\author{E Baracchini$^{1,2}$,
L Benussi$^{3}$,
S Bianco$^{3}$,
C Capoccia$^{3}$, 
M Caponero$^{3,4}$,
G Cavoto$^{5,6}$,
A Cortez$^{1,2}$,
I A. Costa$^{7}$,
E Di Marco$^{5}$,
G D'Imperio$^{5}$,
G Dho$^{1,2}$,
F Iacoangeli$^{5}$,
G Maccarrone$^{3}$,
M Marafini$^{5,8}$,
G Mazzitelli$^{3}$,
A Messina$^{5,6}$,
R A. Nobrega$^{7}$,
A Orlandi$^{3}$,
E Paoletti$^{3}$,
L Passamonti$^{3}$,
F Petrucci$^{9,5}$,
D Piccolo$^{3}$,
D Pierluigi$^{3}$,
D Pinci$^{5}$,
F Renga$^{5}$,
F Rosatelli$^{3}$,
A Russo$^{3}$,
G Saviano$^{3,10}$
R Tesauro$^{3}$,
and S Tomassini$^{3}$}

\address{$^1$Gran~Sasso~Science~Institute, L'Aquila, Italy}
\address{$^2$INFN, Laboratori Nazionali del Gran Sasso, Assergi, Italy}
\address{$^3$INFN, Laboratori Nazionali di Frascati, Frascati, Italy}
\address{$^4$ENEA Centro Ricerche Frascati, Frascati, Italy}
\address{$^5$INFN, Sezione di Roma, Roma, Italy}
\address{$^6$Dipartimento di Fisica, Sapienza Universit\`a di Roma, Roma, Italy}
\address{$^7$Universidade Federal de Juiz de Fora, Juiz de Fora, Brasil}
\address{$^8$Museo Storico della Fisica e Centro Studi e Ricerche ``Enrico Fermi'', Roma, Italy}
\address{$^9$Dipartimento di Matematica e Fisica, Universit\`a Roma TRE, Roma, Italy}
\address{$^{10}$Dipartimento di Ingegneria Chimica, Materiali e Ambiente, Sapienza Universit\`a di Roma, Roma, Italy}
\ead{emanuele.dimarco@roma1.infn.it}

\vspace{10pt}
\begin{indented}
\item[]24 July 2020
\end{indented}

\begin{abstract}
The search for a novel technology, able to detect and reconstruct
nuclear recoil events in the keV energy range, has become more and
more important now that vast regions of high mass WIMP-like Dark
Matter candidate have been excluded.  Gaseous Time Projection Chambers
(TPC) with optical readout are very promising candidates combining the
complete event information provided by the TPC technique with the high
sensitivity and granularity of latest generation light sensors.  A TPC
with an amplification at the anode, obtained with Gas Electron
Multipliers (GEM), was tested at the Laboratori Nazionali di
Frascati. Photons and neutrons from radioactive sources were employed
to induce recoiling nuclei and electrons with kinetic energy in the
range 1--100 \keV.  A He-CF$_4$ (60/40) gas mixture was used at
atmospheric pressure and the light produced during the multiplication
in the GEM channels was acquired by a high position resolution and low
noise CMOS camera and a photomultiplier. A multi-stage pattern
recognition algorithm based on an advanced clustering technique is
presented here. A number of cluster shape observables are used to
identify nuclear recoils induced by neutrons originated from a \ambe
source against X-ray \fe photo-electrons. An efficiency of 18\% to
detect nuclear recoils with an energy of about 6 keV is reached, while
suppressing 96\% of the \fe photoelectrons, making this optically read
out gas TPC a very promising candidate for future investigations of
ultra-rare events as directional direct Dark Matter searches.

\end{abstract}

\section{Introduction}

The advent of a market of single-photon light sensors with high
position resolution can open a new opportunity to investigate
ultra-low rate phenomena as Dark Matter (DM) particle scattering on
nuclei in a gaseous target.

The nature of DM is still one of the key issues to understand our
Universe \cite{PhysRevLett.39.165,Undagoitia_2015}.  Different models
predict the existence of neutral particles with a mass of few GeV or
higher that would fill our Galaxy
\cite{PhysRevLett.113.171301,PhysRevD.79.115016,doi:10.1142/S0217751X13300287,ZUREK201491}. They
could interact with the nuclei present in ordinary matter producing
highly ionizing nuclear recoils but with a kinetic energy as small as
few keV. Moreover, given the motion of the Sun in the Milky Way
towards the Cygnus constellation, such nuclear recoils would exhibit
in galactic coordinates a dipole angular distribution in a terrestrial
detector \cite{MAYET20161}.  In this paper the use of a current
generation CMOS camera providing very low noise level together with a
large granularity (often indicated as {\it scientific CMOS} or sCMOS)
to capture the light emitted by Gas Electron Multipliers (GEMs) in a
Time Projection Chamber (TPC) device is described. The GEMs are
located in the TPC gas volume at the anode position and are used to
amplify the ionization produced in the gas by the nuclear recoils and
other particles. A secondary scintillation light is produced in the
amplification of the avalanche process by the GEM.  This light and its
spatial distribution are reconstructed in the detector and
characterized by means of a clustering algorithm.

Different type of particles will produce distinctive and diverse
patterns of ionization charge, and therefore of light emitted by the
GEMs, given the different way they deposit energy and interact with
matter.  Therefore, nuclear recoils can be efficiently identified and
separated from different kinds of background down to a few \keV
kinetic energy.  The study of the optical readout of a TPC has been
recently conducted with several small size prototypes
(NITEC~\cite{JINST:nitec},
ORANGE~\cite{NIM:Marafinietal,bib:jinst_orange2},
\lemon~\cite{bib:eps, bib:ieee17,bib:elba}) with various particle
sources, in the context of the \cygno
project~\cite{Abritta_Costa_2020,CYGNOweb}, as well as for other
experiments, as ARIADNE~\cite{Roberts_2019}. In the following, the
study of nuclear recoils excited by neutrons from an \ambe source and
electron recoils from a \fe source in the gas volume of the \lemon
prototype is presented.

\section{Experimental layout}
\label{sec:layout}
\begin{figure}[ht]
	\centering
	\includegraphics[width=0.90\linewidth]{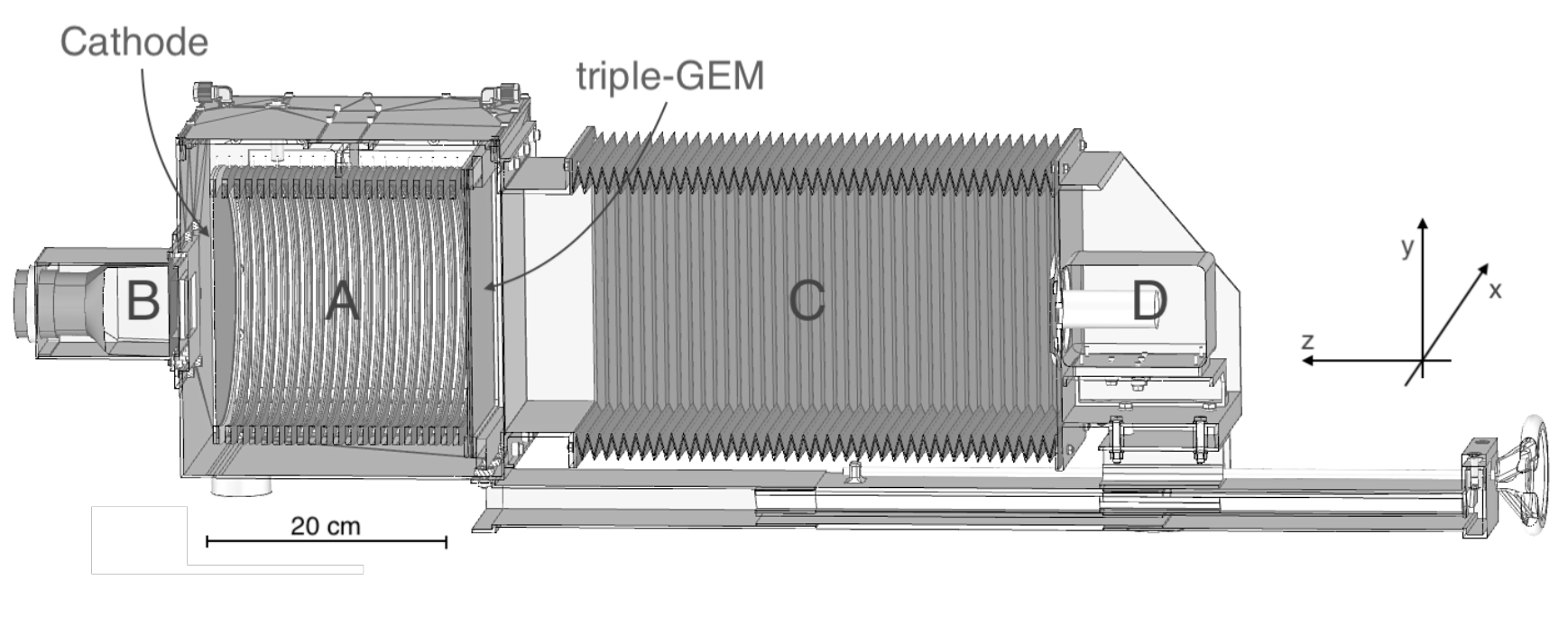} \\
        \includegraphics[width=0.30\linewidth]{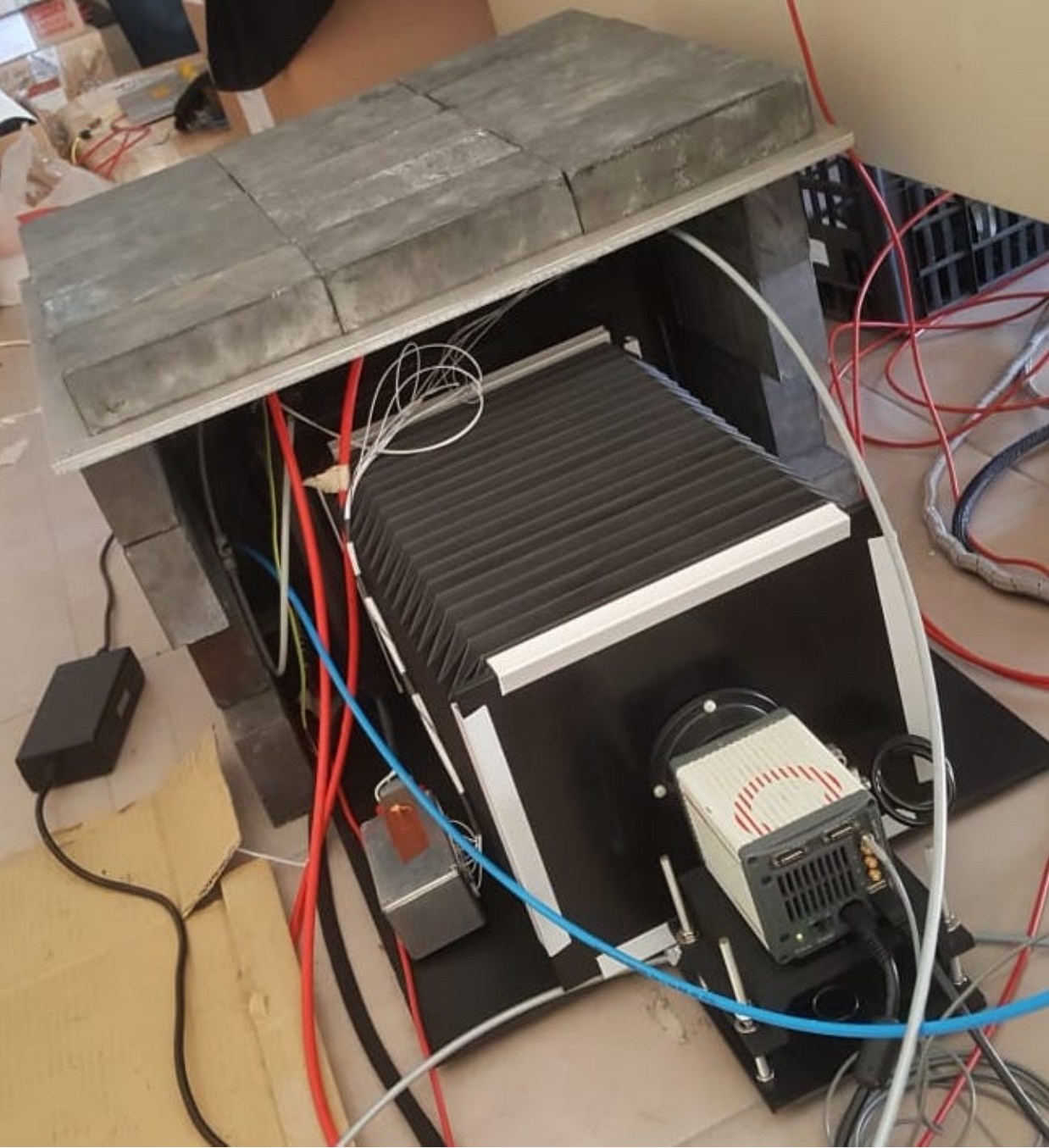}
  	\caption{Top: the \lemon prototype with  its 7 liter sensitive
          volume (A), the PMT (B), the adjustable bellows (C) and the
          sCMOS camera with its lens (D). Bottom: \lemon
          with the lead shield around the drift volume cage. The sCMOS
          camera (on the front) is looking at the GEMs through a
          blackened bellows.
  	\label{fig:lemon}}
\end{figure}

A 7 liter active sensitive volume TPC (named \lemon) was employed to
detect the particle recoils. A sketch (not to scale) of the detector
setup is shown in Fig.~\ref{fig:lemon} (left), while an image of the
detector in the experimental area is shown in Fig.~\ref{fig:lemon}
(right).
The sensitive volume, where the ionization electrons are drifting,
features a 200$\times$240\unit{mm$^2$} elliptical field cage with a
200 mm distance between the anode and the cathode. The anode side is
instrumented with a $200{\times}240$\unit{mm$^2$} rectangular triple
GEM structure.  Standard LHCb-like GEMs (70\unit{$\mu$m} diameter
holes and 140\unit{$\mu$m} pitch)~\cite{bib:thesis} were used with two
2\unit{mm} wide transfer gaps between them. The light emitted from the
GEMs is detected with an \texttt{ORCA-Flash~4.0} camera \cite{ORCAcamera}
through a $203{\times}254{\times}1$\unit{mm$^3$} transparent window
and a bellows of adjustable length.  This camera is positioned at a 52
cm distance from the outermost GEM layer and is based on a sCMOS
sensor with high granularity ($2048{\times}2048$ pixels), very low
noise (around two photons per pixel), high sensitivity (70\% quantum
efficiency at 600\unit{nm}) and good
linearity~\cite{bib:jinst_orange1}. This camera is instrumented with a
Schneider lens, characterized by an aperture f/0.95 and a focal length
of 25\unit{mm}. The lens is placed at a distance $d=50.6$\unit{cm}
from the last GEM in order to obtain a de-magnification $\Delta =
(d/f) - 1 = 19.25$ to image a surface $25.6{\times}25.6$\unit{cm$^2$}
onto the $1.33{\times}1.33$\unit{cm$^2$} sensor.  In this
configuration, each pixel is therefore imaging an effective area of
$125{\times}125$\unit{$\mu$m$^2$} of the GEM layer. The fraction of
the light collected by the lens is evaluated
to be $1.7{\times}10^{-4}$~\cite{bib:jinst_orange1}.
A semi-transparent mesh was used as a cathode in order to collect
light on that side also with a $50{\times}50$\unit{mm$^2$}
\texttt{HZC~Photonics~XP3392} photomultiplier~\cite{PMTPhotonics}
(PMT) detecting light through a transparent
$50{\times}50{\times}4$\unit{mm$^3$} fused silica window. More details
on the \lemon detector can be found in Ref.~\cite{paperBTF}.

A 5\unit{cm} thick lead shielding was mounted around the \lemon field
cage to reduce the environmental natural radioactivity
background. From the measurements of the GEM current with and without
the lead shielding, a factor two reduction in the total ionization
within the sensitive volume, very likely due to environmental
radioactivity, was estimated.

\section{Particle images in the \lemon gas volume}

The \lemon detector was operated in an overground location at
Laboratori Nazionali di Frascati (LNF) with a He-CF$_4$ (60/40) gas
mixture at atmospheric pressure. The triple GEM system was operated
with a voltage across each GEM sides of 460\unit{V} and with a
transfer field between the GEM layers of 2.5\unit{kV/cm}.  A
\texttt{CAEN} \texttt {SY1527} high-voltage module with six
independent channels ensured an adequate voltage stability, and
monitored the bias currents with a precision of 20\unit{nA}. The gas
mixture was kept under a continuous flow of about 200\unit{cc/min} and
with the GEMs operated at a $2.0\times10^6$ electric gain. The typical
photon yield for this type of gas mixtures has been measured to be
around 0.07 photons per avalanche electron\cite{bib:jinst_orange1,
  bib:roby, bib:tesinatalia}, and therefore the overall light gain is
about $10^6$. The field cage was powered by a
\texttt{CAEN~N1570}~\cite{CAENN1570}, generating an electric field of
0.5 \unit{kV/cm}.
  
The motion of particles within the gas mixtures was studied by means
of different simulation tools. In particular, the 
\garfield~\cite{bib:garfield,bib:garfield1} program was used to
evaluate the transport properties for ionization electrons in the
sensitive volume for an electric field of  0.5 \unit{kV/cm}.

Given the diffusion in the gas, ionization electrons produced at a
distance $z$ from the GEM will distribute over a region on the GEM
surface with  a Gaussian transverse profile having : 
\begin{equation}
\label{eq:diff}
\sigma = \sqrt{\sigma^2_0 +  D^2 \cdot z},
\end{equation}
where $D$ is the transverse diffusion coefficient, whose value at room
temperature 140\unit{$\mu{\mathrm{m}}/\sqrt{\mathrm{cm}}$} was
obtained with this simulation.  The value of $\sigma_0$ was measured to
be about 300\unit{$\mu$m} \cite{bib:btf,bib:fe55New}. Therefore, in
average, a point-like ionization will result in a spot of
3--4\unit{mm$^2$}.

\begin{figure}[ht]
  \begin{center}
    \includegraphics[width=0.49\linewidth]{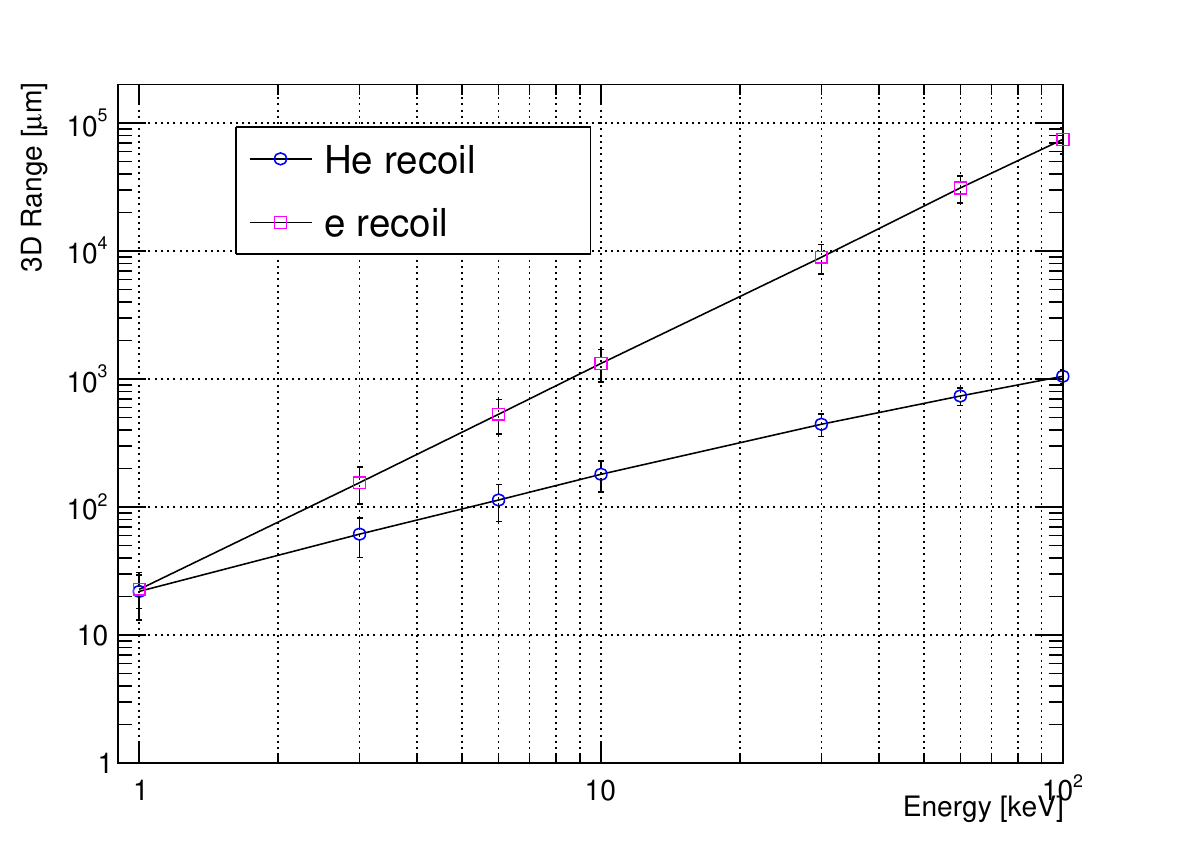}
    \caption{Average ranges for electron and He-nucleus recoils as a
      function of their initial kinetic energy. Lines in the figure are guide for the eye.
      \label{fig:range}}
      \end{center}
\end{figure}

The expected effective ranges of electron and nuclear recoils were
evaluated respectively with \GEANTfour~\cite{GEANT4} and with
\SRIM~\cite{bib:srim} simulation programs. The recoil range estimated
from simulation, as a function of the impinging particle kinetic
energy, is shown in Fig.~\ref{fig:range} for electrons and -as an
example- for He-nuclei.
These results show that:
\begin{itemize}
    \item He-nuclei recoils have a sub-millimeter range up to energies
      of 100\keV and are thus expected to produce bright spots with
      sizes mainly dominated by diffusion;
    \item low energy (less than 10\keV) electron recoils are in
      general longer than He-nucleus recoils with same energy and are
      expected to produce sparse and less intense spot-like signals. For a
      kinetic energy of 10\keV, the electron range becomes longer than
      1\unit{mm} and for few tens of keV, tracks of few cm are
      expected.
\end{itemize}

\begin{figure}[ht]
  \begin{center}
    \includegraphics[width=0.49\linewidth]{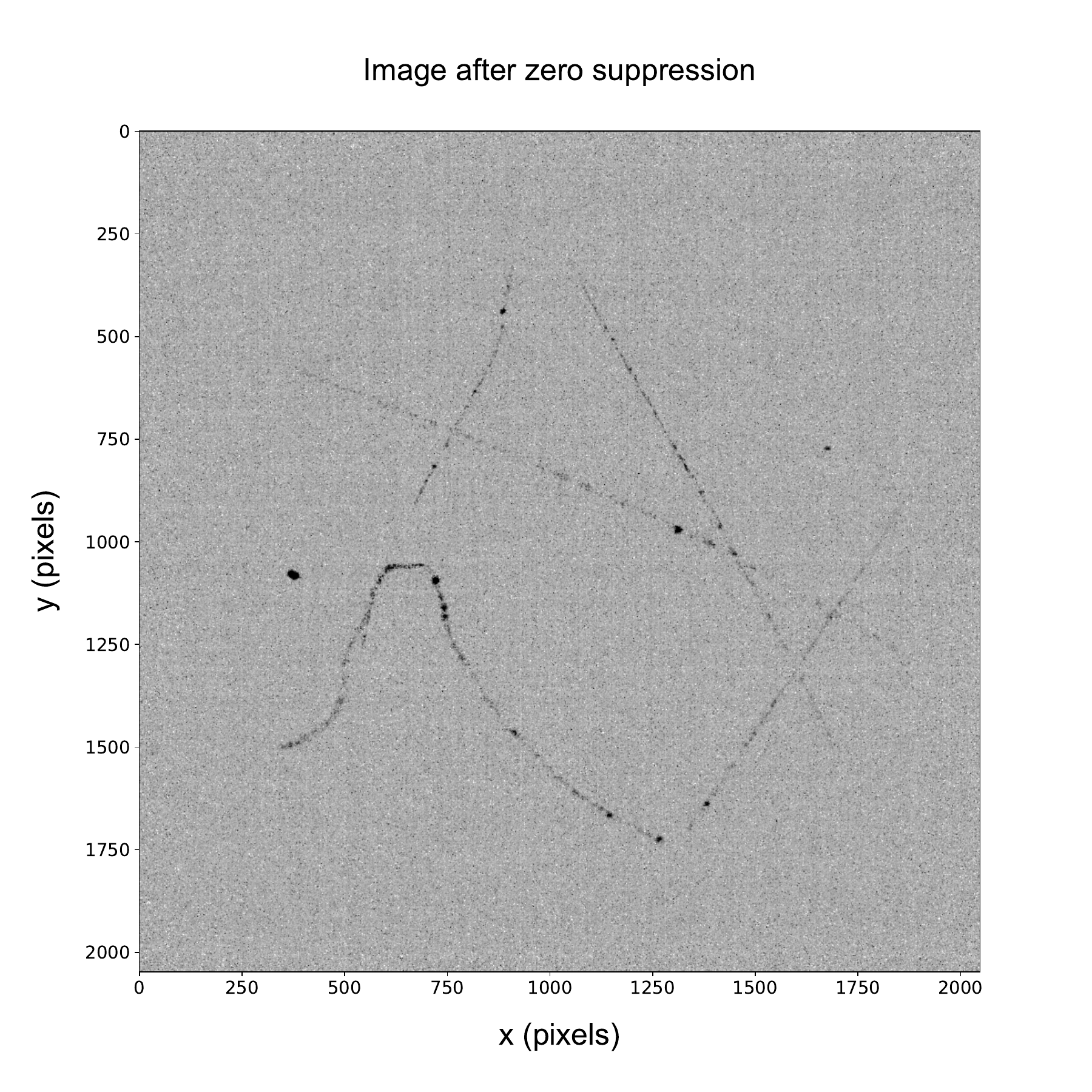}
    \includegraphics[width=0.49\linewidth]{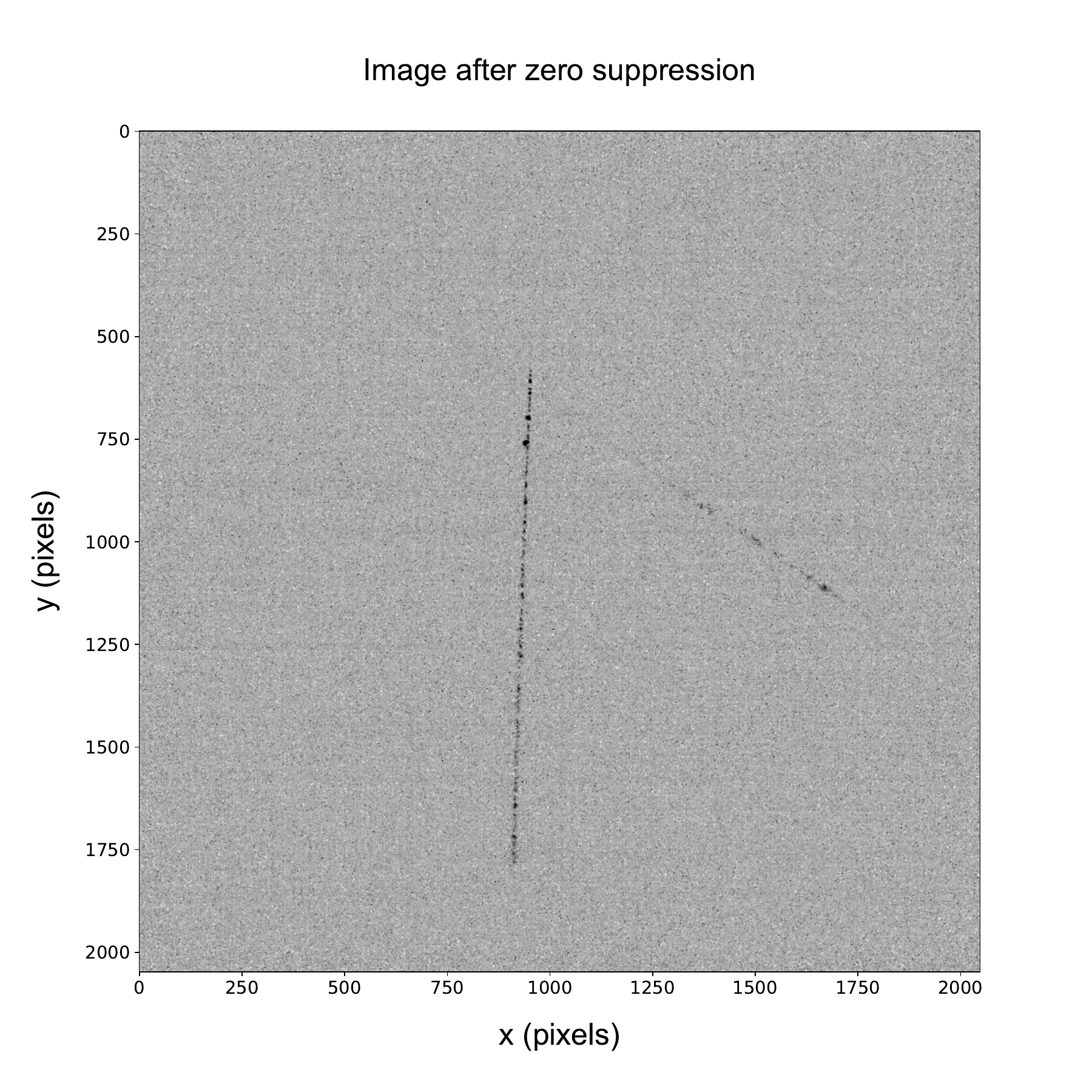}
    \caption{Two example pictures taken with the sCMOS camera with a
      30\unit{ms} exposure time and with a common noise level
      subtracted (zero suppression), belonging to a data taking run
      without any artificial source. Left: cosmic tracks and natural
      radioactivity signals are present. Right: only two long cosmic
      rays tracks are visible. The coordinates are defined such that
      the vertical direction is along the y-axis and cosmic rays are
      expected to come from the top of the figure.
      \label{fig:typicalimage1}}
  \end{center}
\end{figure}

 The images collected by the sCMOS camera contain several instances of
 the particles tracks described above. The sCMOS sensor was operated
 in continuous mode with a global exposure time of
 30\unit{ms}. Example images are shown in
 Fig.~\ref{fig:typicalimage1}.

The PMT waveform was sent to a digitizer board with a sampling
frequency of 4\unit{GS/s}. The trigger scheme of the detector is based
on the PMT signal: if, during the exposure time window, the PMT
waveform exhibits a peak exceeding a threshold of 80\unit{mV}, it is
acquired in a time window of 25\unit{$\mu$s} and the corresponding
sCMOS image is stored.  The digitizer is operated in single-event
mode. No more than one 25\unit{$\mu$s} long PMT waveform is recorded
in each sCMOS exposure time, even if during the sCMOS exposure time
several PMT signals are produced.  Therefore, the PMT information was
mainly exploited only to select frames with a cosmic-ray track.

Several light spots are visible with different ionization patterns due
to different types of particles interacting in the gas.
Figure~\ref{fig:typicalimage1} (left) shows an image with typical long
tracks from cosmic rays traveling through the full gas volume, where
clusters of light with larger energy deposition are clearly visible,
superimposed to low energy electrons, very likely due to natural
radioactivity.  Figure~\ref{fig:typicalimage1} (right) shows an
example of a cleaner event with two straight cosmic ray tracks, that
can be used for energy calibration purposes, since the energy releases
along the path, $dE/dx$, can be predicted given the gas mixture and
pressure.

Two different artificial radioactive sources were employed for testing
and studying the detector responses.

\vspace{10pt}

A {\bf neutron} source, based on a $3.5{\times}10^3$\unit{MBq} activity
$^{241}$Am source contained in a Beryllium capsule (\ambe) was placed
at a distance of 50\unit{cm} from the sensitive volume side.  Because
of the interactions between $\alpha$ particles produced by the
$^{241}$Am and the Beryllium nuclei, the \ambe source isotropically
emits:
 \begin{itemize}
     \item photons with an energy of 59\keV produced by $^{241}$Am;
     \item neutrons with a kinetic energy mainly in a range
       1--10\MeV
     \item photons with an energy of 4.4\MeV produced along with
       neutrons in the interaction between $\alpha$ and Be nucleus.
 \end{itemize}
 The presence of a lead shield around the sensitive volume absorbed
almost completely the 59\keV photon component. A small faction of it
reached the gas through small gaps accidentally present between the lead
bricks.

\vspace{10pt}

A \fe source emitting {\bf X-rays} with a main energy peak at 5.9\keV.
This is the standard candle for calibration and performance evaluation
of \lemon, and its extensive use is documented in
Ref.~\cite{bib:fe55}.

\vspace{10pt}

Four different sets of runs have been recorded: (i) without any source
and no electric signal amplification in the GEMs, to study the sensor
electronic noise; (ii) without any source, but the detector fully
active, to study the ambient background, mainly muons from cosmic rays
and natural radioactivity; runs with either the (iii) \fe source, to
study the detector response to a known signal or with the \ambe
source, to study the \lemon performances in presence of nuclear
recoils.

\vspace{10pt}

\begin{figure}[ht]
  \begin{center}
    \includegraphics[width=0.49\linewidth]{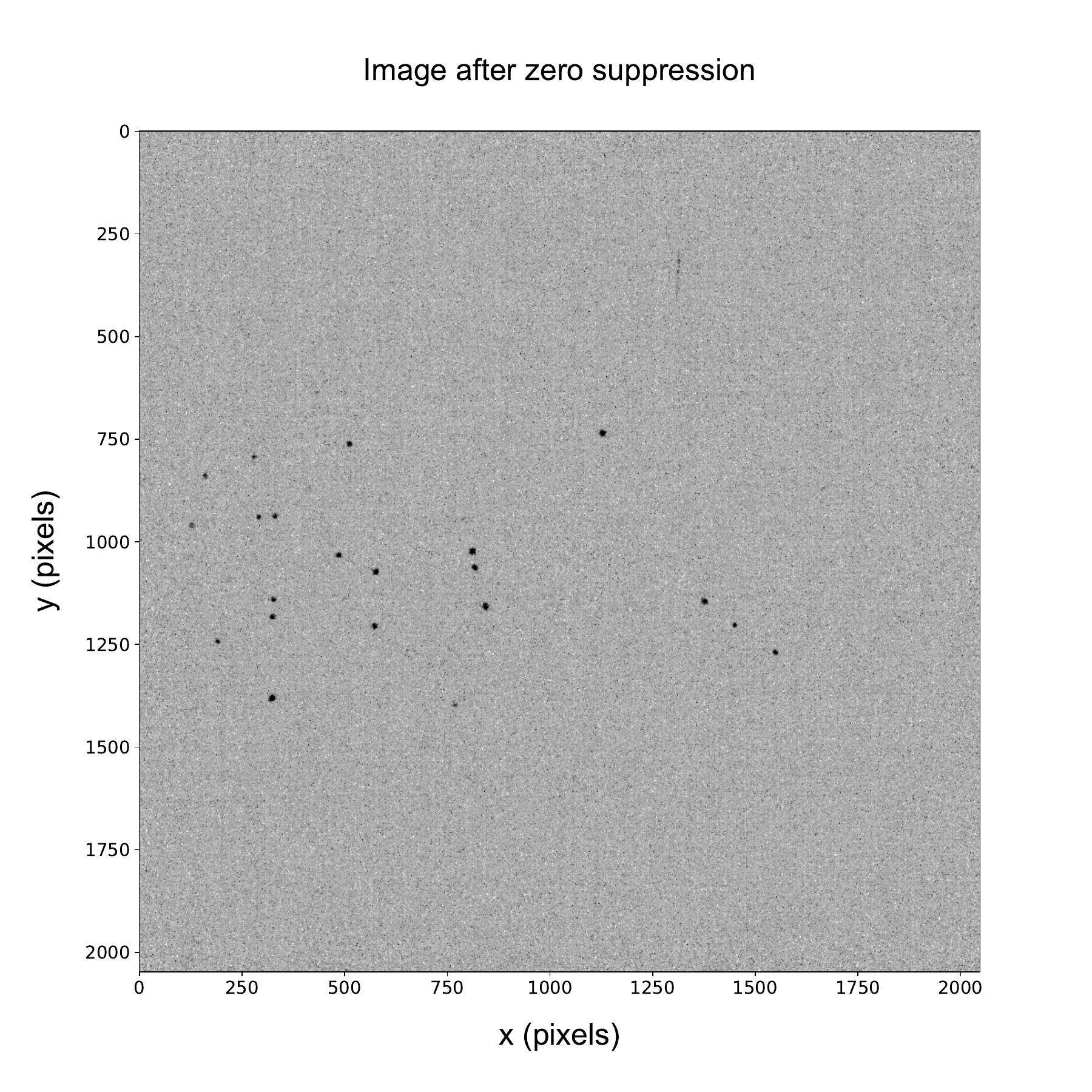}
    \includegraphics[width=0.49\linewidth]{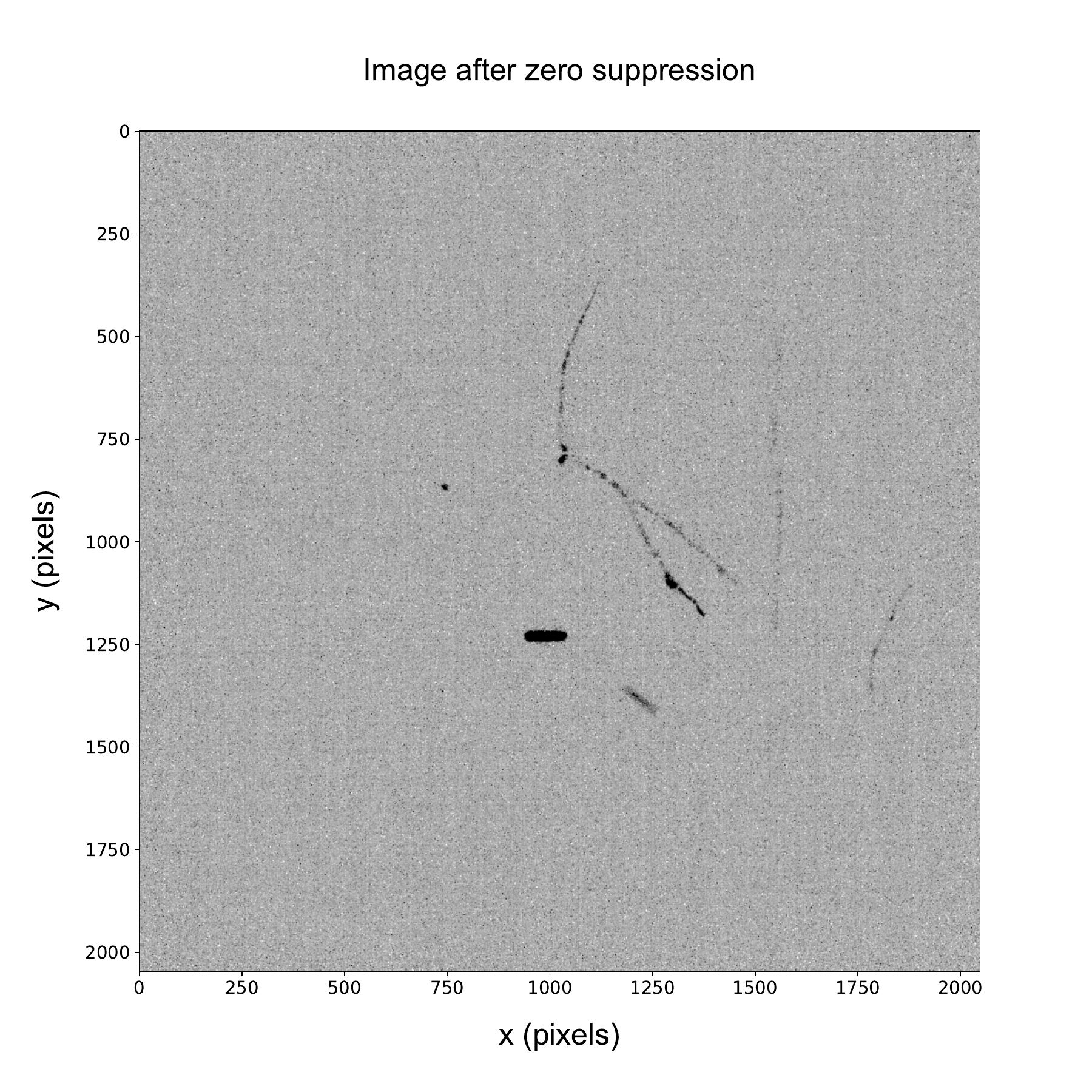}
    \caption{Two pictures taken with the sCMOS camera with a 30 ms
      exposure time. Left: picture taken in presence of \fe
      radioactive source. Right: a nuclear recoil candidate is
      present, in an image with \ambe radioactive source, together
      with signals from natural radioactivity.  The coordinates are
      defined such that the vertical direction is along the
      y-axis.  \label{fig:signals}}
  \end{center}
\end{figure}

Figure~\ref{fig:signals} shows images recorded with the same
30\unit{ms} exposure time, in presence of one of the two sources. The
left panel shows an example of several light spots, characteristic of
energy deposits due to \fe low energy photons.  The right panel shows
a frame recorded in presence of the \ambe radioactive source: the
short and bright track well visible in the center is very likely due
to an energetic nuclear recoil induced by a neutron scattering.

\clearpage
 
\section{Cluster reconstruction algorithm}
\label{sec:clustering}
The light produced in the multiplication process through the GEMs and
detected by the sCMOS sensor is associated in clusters of neighboring
pixels. This is achieved by following the trail of energy deposition
of the particle traveling through the gas of the sensitive volume. The
energy released as ionization electrons is estimated by the amount of
the light collected by the sensor.  In the range from few keV to tens
of keV for stopped electrons, the deposited energy is equivalent to
their total kinetic energy, while for stopped nuclei it represent only
a fraction of their initial kinetic energy.  Therefore, it is of
primary importance to have a reconstruction algorithm that includes
all the camera pixels hit by the real photons originating from the
energy deposits, while rejecting most of the electronic noise from the
camera sensor. Noise can either create fake clusters or, more likely,
add pixels in the periphery of clusters originated by real photons,
thus biasing the energy estimate.  Possible additional noise, arising
for example from GEM stages, was already demonstrated be negligible
\cite{bib:fe55}.

The energy reconstruction follows a three-steps procedure: the
single-pixel noise suppression is briefly described in
Section~\ref{sec:zerosuppression}. This is followed by the proper
clustering: first the algorithm to form basic clusters from single
small deposits is described in Section~\ref{sec:basiccl}, then the
supercluster method, aiming to follow the full particle track, and
seeded by the basic clusters found in the previous step, is described
in Section~\ref{sec:supercl}.

The results of this paper are based on the properties of the
reconstructed superclusters and are described in
Section~\ref{sec:clustershapes}.

\subsection{Sensor noise suppression}
\label{sec:zerosuppression}
The electronic noise of the sensor was estimated in data-taking runs
acquired with the sensor in complete dark, obtained by covering the
camera lens with its own cap or, equivalently, lowering the voltage
across the GEM electrodes to 300\unit{V} ({\it pedestal} runs). The
latter option, which was demostrated to be fully equivalent to the
former, is a valuable method to measure the sensor noise periodically,
and track its evolution, during the periods without data taking of
the \cygno experiment.

\begin{figure}[ht]
  \centering
  \includegraphics[width=0.45\linewidth]{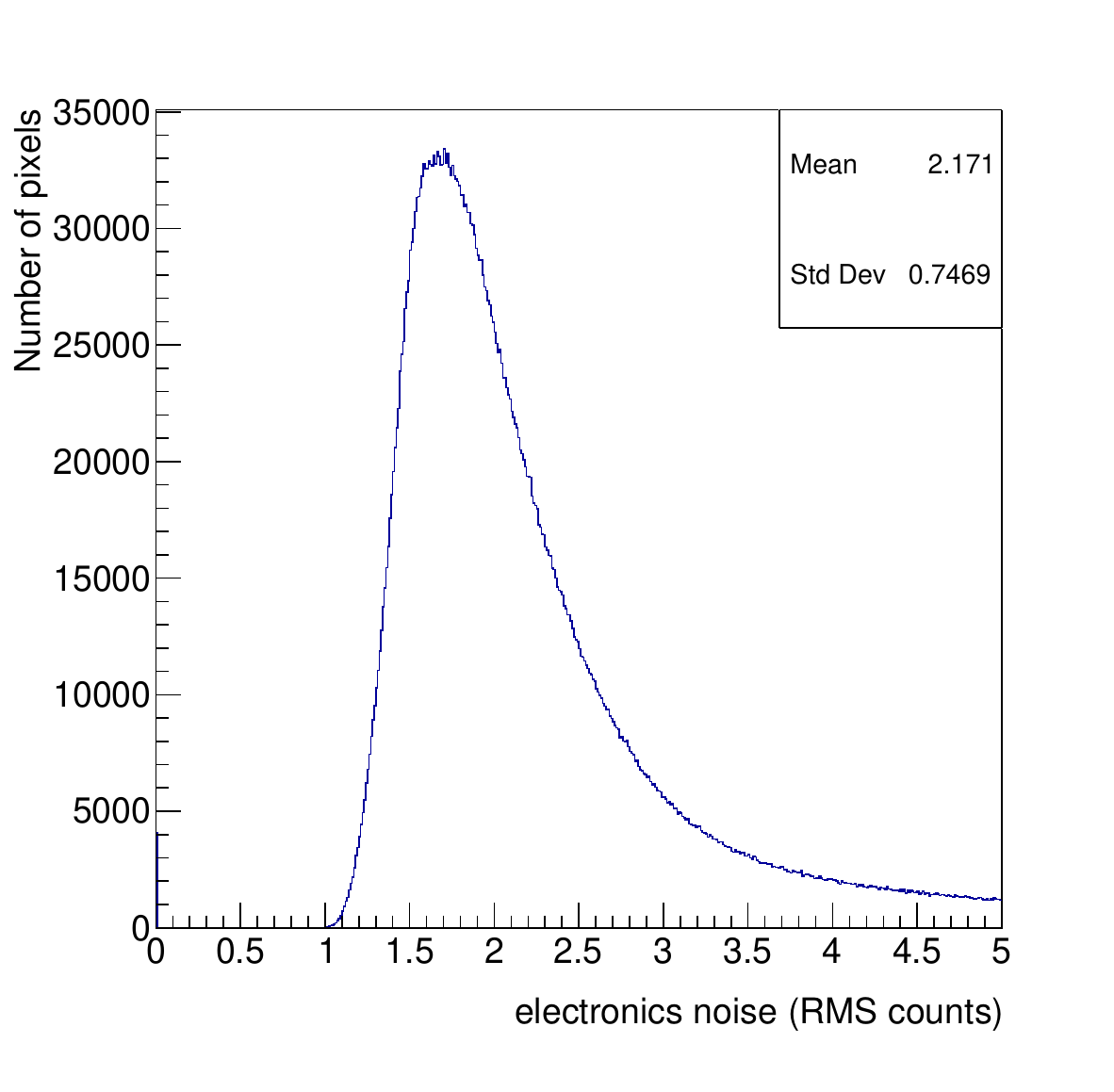}
  \caption{Distribution of the electronic noise of the sensor,
    estimated in images taken with sensor in complete dark, and
    evaluated as the SD of the distribution of the counts for each
    pixel.  \label{fig:noise}}
\end{figure}

For each pixel, the pedestal was computed as the average of the counts
over many frames, while the electronic noise was estimated as their
standard deviation (SD). The distribution of the pixels SD is shown in
Fig.~\ref{fig:noise}. The mode of this distribution is about 1.8
photons per pixel, but a tail is present, with pixels having a noise
of more than 5 photons per pixel. For such pixels, a very non-Gaussian
distribution was observed, while for the pixels in the bulk of the
distribution, the pedestal distribution followed a Gaussian shape. To
form the pedestal-subtracted image, the pedestal mean $\mu_i$ was
subtracted to the image for each $i^{th}$ pixel, to account for the
non-uniformity of the pedestal mean across the sensor. An initial
noise suppression was applied by neglecting the pixels with counts
less than $1.3\,\textrm{SD}_i$.
On such pedestal-subtracted zero-suppressed images an upper threshold
was applied to reject hot pixels, which are more likely due to sensor
instabilities than to a real energy release. They are found to be not
malfunctioning pixels since this behavior disappears after a power
cycle of the camera and their response to light becomes homogeneous to
the neighbors: therefore a dynamic (run-by-run) suppression is needed.
They are efficiently identified as high-intensity, isolated pixels,
and distinguished by a true energy deposit, for which each pixel is
surrounded by some other active pixels. A threshold is applied on the
ratio $R_9$ between the pixel and the average of the counts in a
$3{\times}3$ pixels matrix surrounding it, and a minimum number of two
pixels above noise in that matrix is required to discriminate good
pixels from hot ones. Only good pixels are retained for the subsequent
clustering.

The resolution of the resulting image is initially reduced by forming
\textit{macro-pixels}, by averaging the counts in $4{\times}4$ pixel
matrices. This is needed to reduce the combinatorics of the subsequent
clustering algorithm, in order to be executed in a reasonable time for
each image. On such $512{\times}512$ pixel map, a median
filter~\cite{medianfilter} is applied, which is effective in
suppressing the electronics noise fluctuations in a $4{\times}4$ pixel
matrix and it is computationally efficient, as described in more
details in Ref.~\cite{medianfilter_cygno}. The output image is passed
to the basic clustering algorithm, described in the following.

\subsection{Basic cluster reconstruction}
\label{sec:basiccl}
The basic clustering algorithm, called \idbscan and described in
details in Ref.~\cite{iDBSCAN}, represents an improvement of the
neighboring pixels clusters, called \nnc, previously used to study the
performance of the \lemon detector with \fe radioactive
source~\cite{bib:fe55}. It is briefly described also here, since it
represents the seeding for the final clustering algorithm.

The energy deposition in the sensitive volume of the TPC is estimated
from the two-dimensional (2D) projection on the $x$--$y$ axes of the
light emitted in the multiplication process within the GEMs
planes. The pattern shows a large variation, depending on the
interacting particle. For images recorded with the \fe calibration
source, the signature of the typical 5.9\keV photons is a spot of few
mm$^2$, with the exact size depending on the diffusion in the
gas, \ie, on the distance from the anode, along $z$, of the point
where the energy release happens (see Fig.~\ref{fig:signals}
left). Muons from cosmic rays travel across the gas volume and leave a
typical signature of a straight track, shown in
Fig.~\ref{fig:typicalimage1} (right), but with several agglomeration
with larger density along the path. Natural radioactivity shows an
irregular pattern, sometimes curly, with several kinks along the
path. Finally, the signal from nuclear recoils due to neutrons,
originated by the \ambe source, is expected to be spot-like, or to
emerge as short straight tracks with a length smaller than 1\unit{mm}
for energies below 100\keV, as shown by Fig.~\ref{fig:range}.

Their track length and their size is found to depend a lot on the
initial energy of the impinging neutron, and also on the mass of the
recoiling nucleus in the He-CF$_4$ gas mixture utilized in
the \lemon detector.

Thus, the clustering algorithm needs to be flexible enough to
efficiently reconstruct a diverse set of patterns, from small round
spots to long and kinky tracks. A first step of the clustering,
called \textit{seeding}, is used: it focuses in the clustering of
spot-like neighboring pixels.  The method applied for the \lemon
detector is an evolution of the classic \dbscan
algorithm~\cite{dbscan}.  This is a non-parametric, density-based
clustering, which groups together pixels above threshold with many
neighbors, within a circle with a radius $\epsilon$. Its distinctive
characteristics making this method very suitable to the \lemon case is
its ability to label as outliers, and so not to include in the
clusters, pixels that lie isolated in low-density regions, \ie, pixels
from electronic noise of the sensor surviving the zero
suppression. The extension of \dbscan used for \lemon data analysis
consists in including a third dimension to the phase space of the
points considered, adding to the pixel position ($x$--$y$ coordinates)
the measured number of photons in that pixel, $N_{ph}$.  This approach
improves the combinatorial background rejection and the energy
resolution with respect the previously used \nnc algorithm, as
described in details in Ref.~\cite{iDBSCAN}.

To be as inclusive as possible, and since different interactions may
have vastly different intensities, even varying along the track, the
clustering procedure is iterated three times.  First, the \dbscan
parameters were tuned to form clusters of dense (in $x$--$y$ dimension)
and intense (in the $N_{ph}$ dimension) pixels. The density in 3D is
called \textit{sparsity}.  This step typically identifies either rare
hot spots of the GEMs, or, efficiently, short nuclear recoils. The
pixels belonging to the reconstructed clusters are then removed from
the image, and the \dbscan procedure is repeated, with looser sparsity
parameters. The second iteration is tuned to efficiently reconstruct
\fe round spots and slices of tracks from nuclear recoils with lower
intensity. It also collects the agglomeration with larger density
along cosmic tracks, clearly visible in the example in
Fig.~\ref{fig:typicalimage1} (right).  A third iteration of \dbscan
with even looser parameters is finally executed, targeting faint
portions of a cluster. These are especially used as a proxy for the
characterization of clustered noisy pixels, while the first two are
used as seeds for the final clustering step, described in
Sec.~\ref{sec:supercl}.

To be computationally viable, the \idbscan basic clustering is
performed on the image with reduced resolution, 512$\times$512. In
typical images this allows the basic cluster reconstruction to be run
in approximately 1\unit{s} on an \texttt{Intel Xeon
E5-2620}~2.00\unit{GHz} and 64\unit{GB} RAM. The reconstruction
algorithm is implemented in \PYTHONthree~\cite{python3}, and
interfaced with CERN \ROOT v6~\cite{root}.

\begin{figure}[ht]
  \begin{center}
     \includegraphics[width=0.49\linewidth]{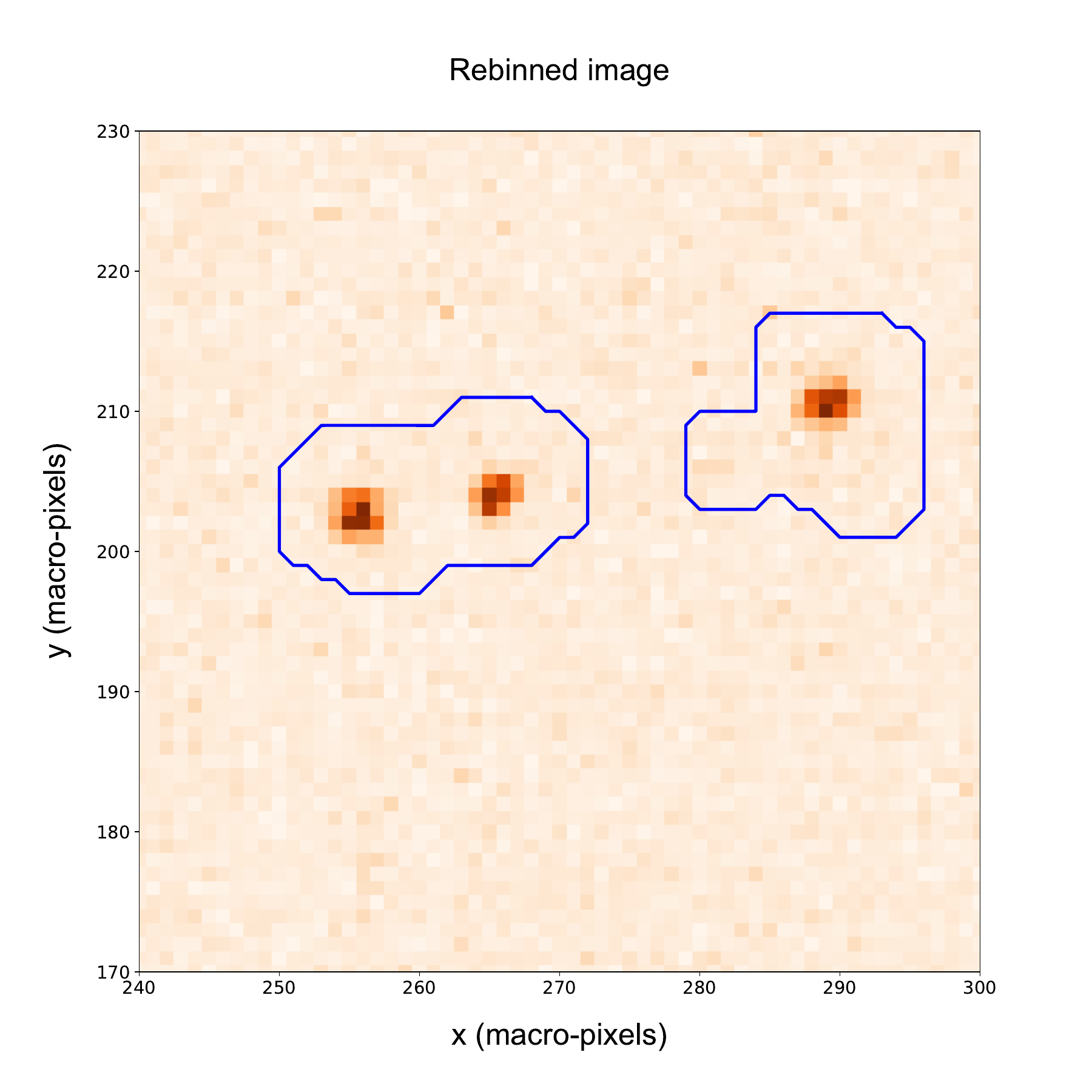}
      \includegraphics[width=0.49\linewidth]{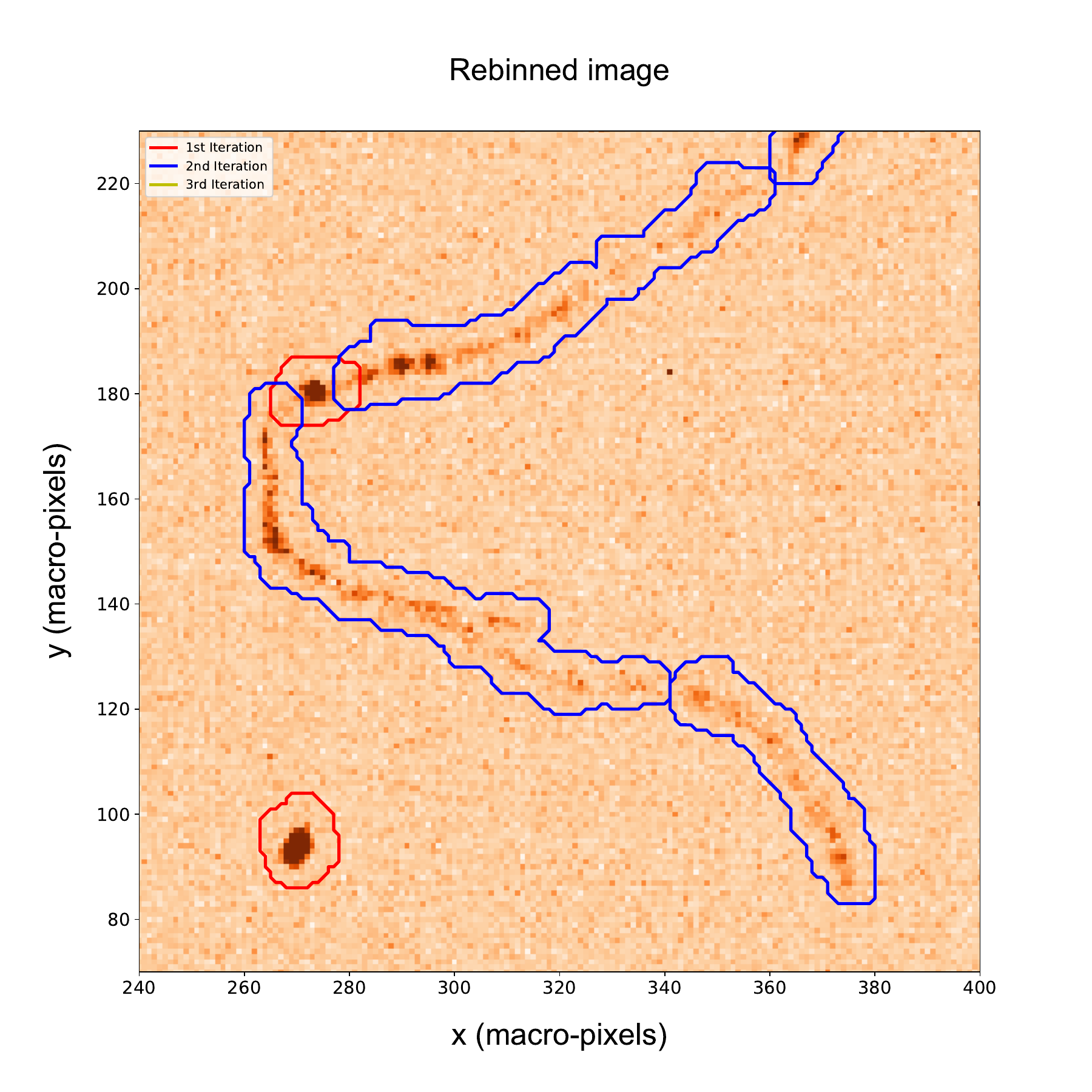}

      \caption{Basic clusters reconstructed with the \idbscan
    algorithm in the low resolution (512$\times$512) image for two
    example events with very different patterns. Continuous lines
    represent the approximate contours of the reconstructed basic
    clusters of the first (red line) or second (blue line) \idbscan
    iteration. Left: clusters on spots from \fe source, two of which
    are merged together. Right: Track from natural radioactivity and a
    nuclear recoil candidate in an event with \ambe source. The long
    track is split in several basic clusters of different \idbscan
    iteration. \label{fig:basic_clusters}}

   \end{center}
\end{figure}

Examples of clustered pixels in two cases are shown in
Fig.~\ref{fig:basic_clusters}. The left panel shows an example of
clusters reconstructed on the low-resolution image of one event
with \fe source. Three spots are clearly visible: one, as typical for
events with this calibration source with a moderate activity, is
reconstructed by a single cluster of the second iteration. The other
two are close enough that are merged in a single cluster of the same
iteration. The cases of merged spots containing twice the energy of a
single X-ray deposit, given the activity of the \fe source, represent
about one tenth of the clusters in this set of runs. The energy
resolution is good enough to distinguish statistically the single and
merged spots, as will be described in Sec.~\ref{sec:calibration}.  The
optimization of the \idbscan parameters is done assuming a low pileup
of events, typical of the running conditions for a future underground
run of the \cygno project, of which \lemon is the prototype, where the
occurrences of such cases are expected to be negligible.

The right panel shows the outcome of the \idbscan algorithm on a
longer track, presumably from natural radioactivity, and one possible
short nuclear recoil.  The nuclear recoil candidate is very dense,
highly-energetic, and isolated. Therefore, it is reconstructed as a
single cluster in the first iteration. The long track shows several
clusters with higher intensity. One of them has a large energy, and it
is reconstructed as an isolated single iteration-1 cluster. The rest
of the track is reconstructed by multiple iteration-2 clusters, which
are split where the energy deposition has a minimum extending across
too many pixels to be joined together in the same cluster. Events like
these, which are frequent for muons, natural radioactivity, but also
signals from $\alpha$-particles with higher energy, originating by the
possible interaction of neutrons with the plastic material of the
field cage, justify the need of the subsequent step of
the \textit{superclustering}, which follows the track without
splitting it in parts. This is described in the following section.

\subsection{Supercluster reconstruction}
\label{sec:supercl}
The aim of the superclustering procedure is to collect the majority of
the pixels belonging to a track which is long and can be split in
multiple parts in the clustering step described before.  Indeed, the
main limitation of \idbscan to follow a long track is mainly
originated by the non uniform energy release along the path length.
As can be clearly seen in Fig.~\ref{fig:basic_clusters} (right), or
even in the example of a raw image of an event with two long cosmic
rays in Fig.~\ref{fig:typicalimage1} (right), clusters with larger
energy release are followed by regions along the path with a lower or
even a zero release.  These local minima are sometimes as large, in
the 2D space, as the typical size of the $\epsilon$ parameter of
\dbscan~\cite{dbscan}. Despite the low electronic noise of the
\texttt{ORCA-Flash 4.0} camera sensor, the energy releases in these local
minima are similar in magnitude to the average single-pixel noise.

The \idbscan is limited in connecting the full length of an extended
path, because of two reasons. First, inflating $\epsilon$ parameter as
much as needed to cover the areas of local minima conflicts with the
need to reject noise around the cluster.  The basic cluster parameters
were optimized for the \lemon running conditions to collect most of
the signals with an energy as low as few keVs and to reject the
typical noise of $\approx 1$ photon per pixel. This avoids collecting
extra noise in the cluster, biasing the energy scale and worsening its
resolution, and keeps the rate of fake clusters at a negligible
level~\cite{iDBSCAN}.  Second, the iterative nature of the algorithm,
with different parameters for each iteration, each tuned for very
different intensity, makes it convenient and efficient for a
deposition of a fixed energy density (like the spots originating from
the \fe source), but not for the cases as in
Fig.~\ref{fig:basic_clusters} (right), where the same track is split
in several parts, some of them belonging to different iterations.
This requires a method that can continuously follow the pattern of the
track, profiting of the full resolution image, where the {\it
gradients} of the energy deposition along the track trajectory are
smaller than the ones in the transverse direction, but still give
information on the energy release pattern. Several existing algorithms
were tested to profit of this, but executing any of them on the full
$2048{\times}2048$ image is not manageable CPU-wise, due to the huge
pixel combinatorics.

Therefore, the procedure adopted for the final supercluster
reconstruction in the
\lemon detector starts from defining the \textit{interesting regions}
in the image that may contain pixels from an energy deposit. These are
identified by the basic cluster algorithm \idbscan previously
described, which is applied on the $512{\times}512$ reduced-resolution
image. In order to gather the peripheral pixels, especially along the
track trajectory where breaks into small basic clusters may have
happened, a window of $5{\times}5$ pixels is considered, around each
pixel belonging to a macro-pixel clustered in a basic cluster. A full
resolution image formed only by the interesting pixels passing the
simple initial filtering described in Sec.~\ref{sec:zerosuppression}
is created.  The gradients of the intensity $N_{ph}$ in such image are
computed pixel-by-pixel to look for the edge region where the image
turns from signal to noise-only:
\begin{equation}
\label{eq:gradient}
\vert\vert\nabla(N_{ph})\vert\vert =
\sqrt{\left(\frac{\partial N_{ph}}{\partial x}\right)^2
  +\left(\frac{\partial N_{ph}}{\partial y}\right)^2},
\end{equation}
while the gradient direction is given by:
\begin{equation}
  \label{eq:graddir}
  \theta = \tan^{-1}\left(\frac{\partial N_{ph}}{\partial y}/\frac{\partial N_{ph}}{\partial x}\right).
\end{equation}
In order to reduce the effect of the noise, which induces fluctuations
in the first derivatives of Eq.~\ref{eq:gradient}, a Gaussian filter
is applied, which smoothen the response by convolving the pixel
intensity with a Gaussian function, having as $\sigma$ the SD of the
intensities of all the pixels considered, and rejecting the ones
falling outside a 5$\sigma$ window.

The superclustering algorithm, applied on the filtered image, is an
application of the \textit{morphological geodesic active
contours}\cite{gac,mgac}, called \gac in the following.  This method
uses an active contour finding, widely used in computer vision, where
the boundary curve $\mathcal{C}$ of an object is detected by
minimizing the \textit{energy} $E$  associated to $\mathcal{C}$:
\begin{equation}
  \label{eq:gacenergy}
  E(\mathcal{C}) = \int_{0}^{1} g(N_{ph})(\mathcal{C}(p)) \cdot \vert\mathcal{C}_p\vert dp,
\end{equation}
where $ds=\vert\mathcal{C}_p\vert dp$ is the arc-length
parameterization of the curve in the 2D space, and $g$ is the stopping
edge function, which allows to select the boundary of the cluster.  In
the \gac method used for the \lemon images, the $g$ function is purely
geometrical, and uses the geodesics of the image, \ie, the local
minimal distance path joining points with the same light intensity
gradient. The function $g(N_{ph})$ is given by:
\begin{equation}
g(N_{ph}) = \frac{1}{\sqrt{1+\alpha\vert\nabla G_\sigma * N_{ph}\vert}},
\end{equation}
which is minimal in the edges of the image.  The $G_\sigma * N_{ph}$ is the
aforementioned $5\sigma$ Gaussian filter, and the parameter $\alpha$,
which regulates the strength of the filter, was tuned on
typical \lemon images to be $\alpha=100$.

This method was chosen because it allows to follow patterns that
may vary from convex to concave shape, eventually with kinks, \eg in
cases of $\delta$-ray emissions. To improve the shrinking of the
cluster boundary in the cases of tracks turning from concave to convex
along their trail, the \textit{balloon} force~\cite{mgac}, which is a
term added to Eq.~\ref{eq:gacenergy} to smooth the cluster contour, is
set to -1, in order to push the contour towards a border in the areas
where the gradient is too small. A number of 300 iterations is used to
evolve the supercluster contour.

\begin{figure}[ht]
  \begin{center}
     \includegraphics[width=0.49\linewidth]{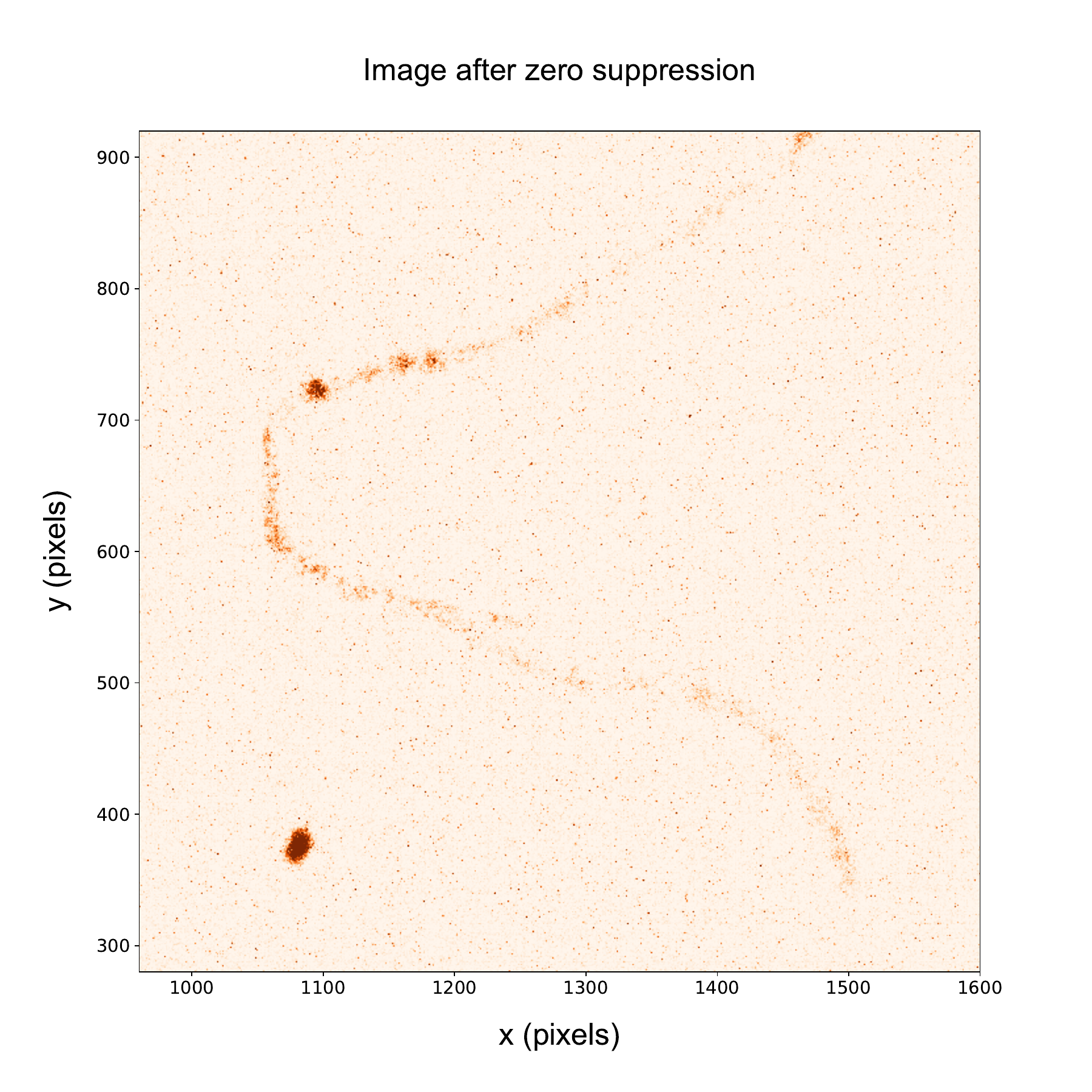}
      \includegraphics[width=0.49\linewidth]{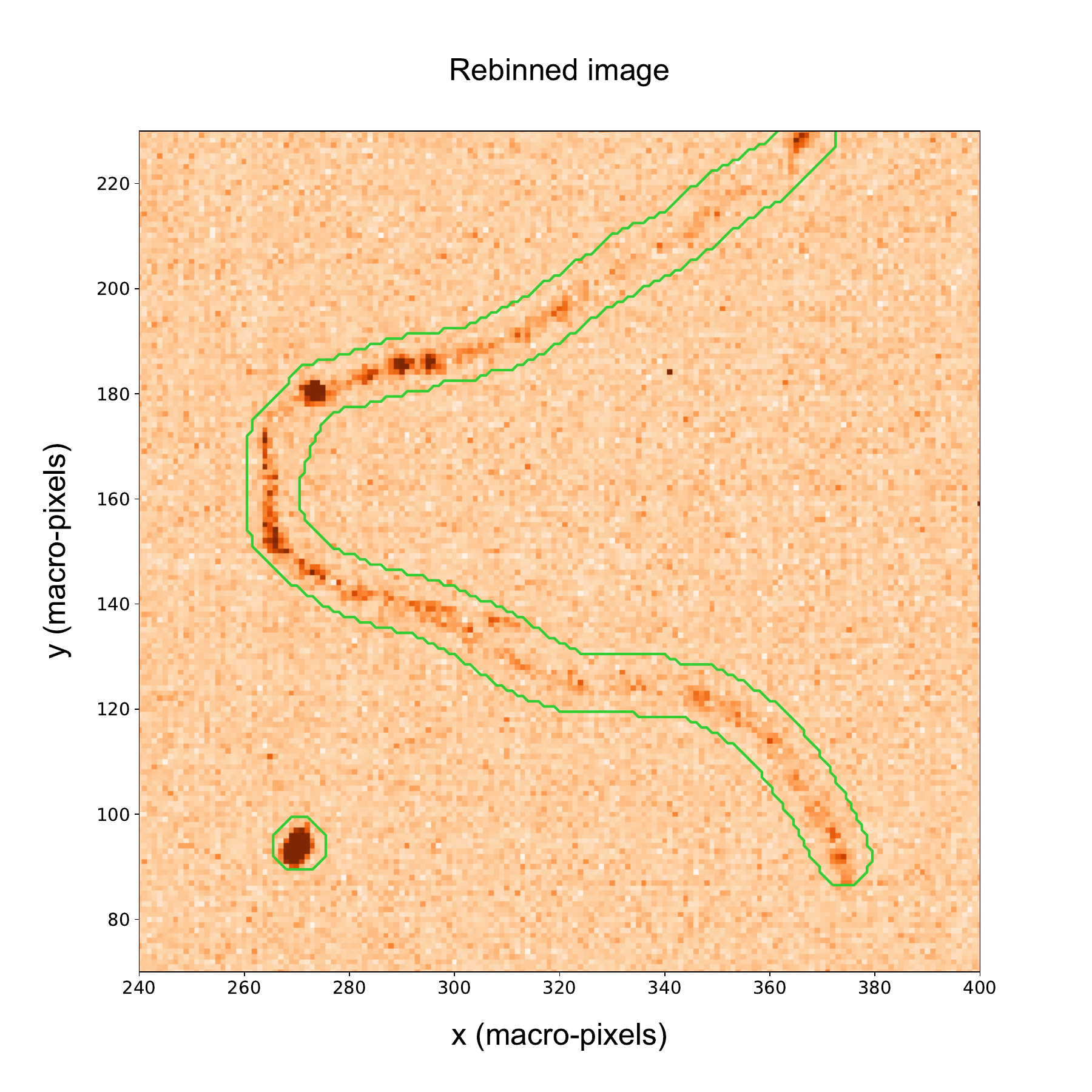}
      \caption{Left: zoom on the full-resolution image of a track
        candidate in a run with the \ambe radioactive source. Right:
        output of the superclustering on the rebinned image. The
        continuous line represents the approximate contour of the
        reconstructed supercluster. \label{fig:super_clusters1}}
  \end{center}
\end{figure}

\begin{figure}[ht]
  \begin{center}
     \includegraphics[width=0.49\linewidth]{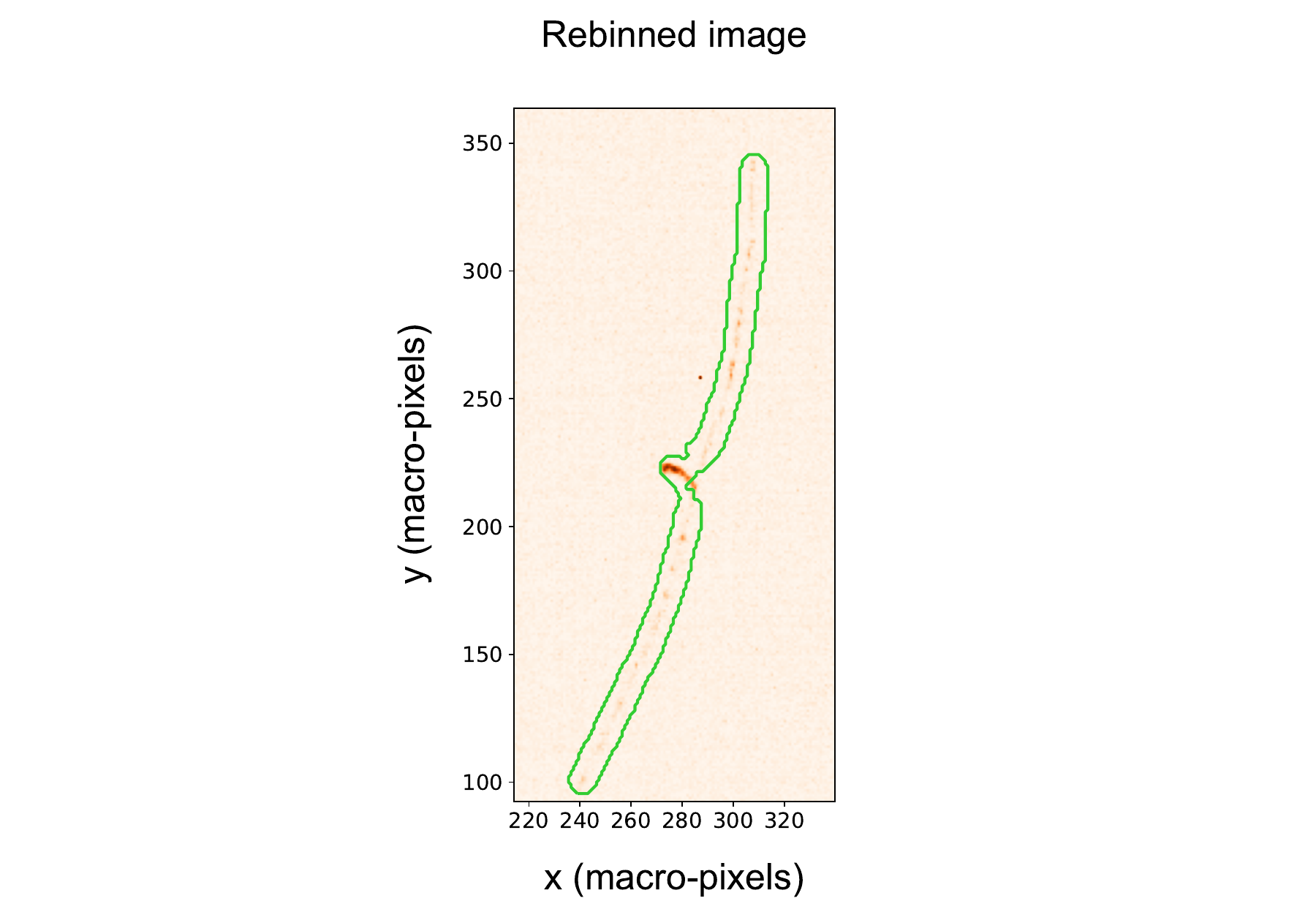}
     \includegraphics[width=0.49\linewidth]{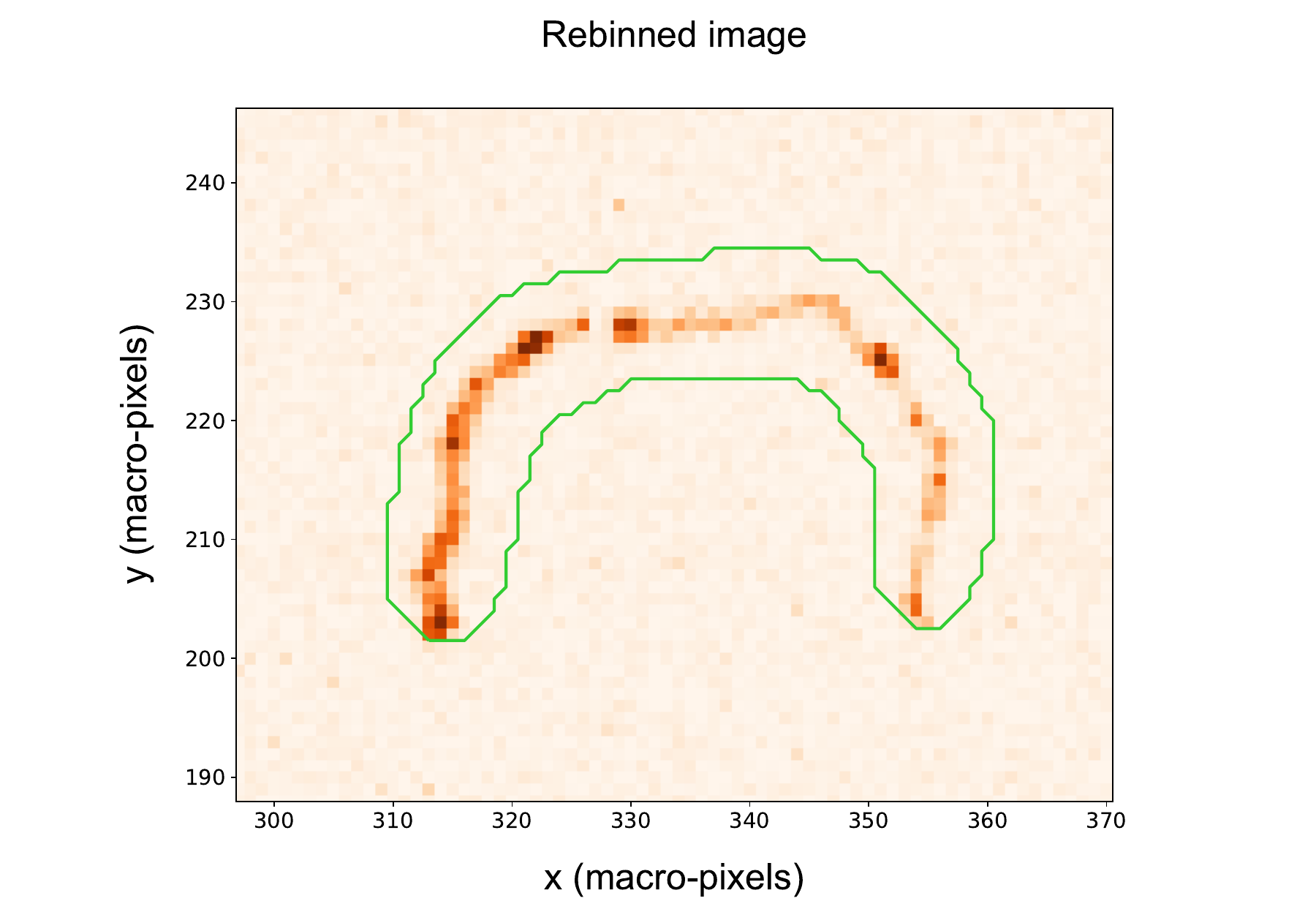} \\
     \includegraphics[width=0.6\linewidth]{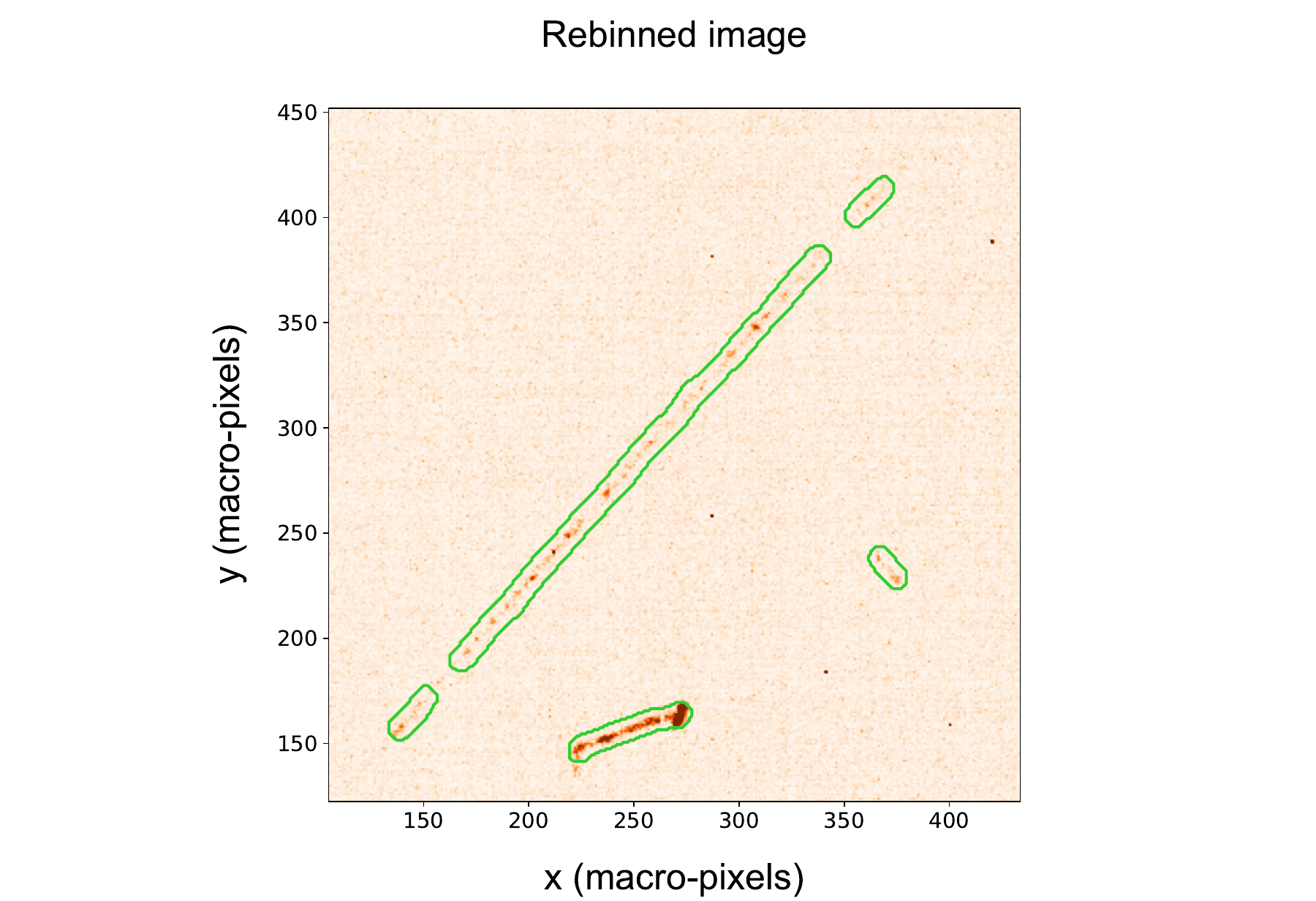}

     \caption{Superclusters reconstructed in a run without artificial
       radioactive sources. The continuous lines represent the
       approximate contours of the reconstructed superclusters. Top
       left: cosmic ray track fully reconstructed by the \gac
       superclustering. A $\delta$-ray is included in the
       supercluster. Top right: curly track from a candidate of
       natural radioactivity interaction. Bottom: a cosmic ray with
       the extremes not joined to the main track, plus a curly track
       from natural radioactivity. \label{fig:super_clusters2}}
       
  \end{center}
\end{figure}

The example track shown in Fig.~\ref{fig:basic_clusters} (right) after
the basic clustering step, is shown again in full resolution, zoomed
around the cluster, in Fig.~\ref{fig:super_clusters1} (left). The
output of the superclustering with the \gac algorithm is shown on the
right panel of the same figure. The splitting of the cluster,
happening at the basic clusters step, is recovered: the portions with
high and low density along the path of the energy release are joined
together. Other three examples of superclustered images are shown in
Fig.~\ref{fig:super_clusters2}, in runs without any artificial
radioactive source. The top left panel shows an example of a cosmic
ray track fully reconstructed by the \gac superclustering, which also
includes a $\delta$-ray in the middle of the track length. The top
right panel shows an example of curly track from a candidate of
natural radioactivity interaction; bottom panel shows an example where
both a cosmic ray and a curly track are present. In this case, the
extremes of the long and straight track are still split, but this is
much rarer than after the basic clustering, and it happens when the
local minima along the trajectories are compatible with noise-only
for more than $\approx$1\unit{cm}.

\clearpage

\subsection{Energy scale calibration using \fe source}
\label{sec:calibration}
The containment of the energy in the supercluster was verified
with simulations of nuclear and electron recoils  within
the gas mixture of the \lemon detector, performed with
\SRIM~\cite{bib:srim}.  For both types of recoils,
for the energy range of interest for DM search, \ie, $E \lesssim
100$\keV, when considering deposits without electronics noise and no
diffusion in the gas, the peak of the $\abs{E-E_{true}}/E_{true}$
distribution is within 5\%. Adding a noise approximated as a Gaussian
function with a mean and a SD equal to the ones observed in the
pedestal runs, and a diffusion following the parameterization in
Eq.~\ref{eq:diff}, the fraction of the true energy contained in the
supercluster decreases to about 80\%.
The decrease in the energy containment in the supercluster is due to
the smearing of the 2D track pattern around the periphery of the
cluster, mostly due to the diffusion effect.  This decreases the
gradients in Eq.~\ref{eq:gradient} around the edges, and so the
supercluster can shrink more around the crest, loosing part of the
tails that can be confused more easily with the noise.  A more
realistic noise description, and an improved diffusion model, based on
the one measured in data is necessary to tune the supercluster
parameters in simulation to recover part of the containment.  The
energy resolution found in simulation (around 4\%) is far from the
measured one in data, around 18\%, because of the absence, in the
simulation, of the dominant contribution of the response fluctuations:
Poissonian distribution of the number of primary electrons ionized in
the gas and exponential behavior of the number of secondary electrons
produced in each GEM amplification stage~\cite{bib:thesis}. Both of
them are expected to give rise to fluctuations of the order to 10\%,
that, once added in quadrature, can account for a large part of the
measured energy resolution.

\begin{figure}[ht]
  \begin{center}
    \includegraphics[width=0.49\linewidth]{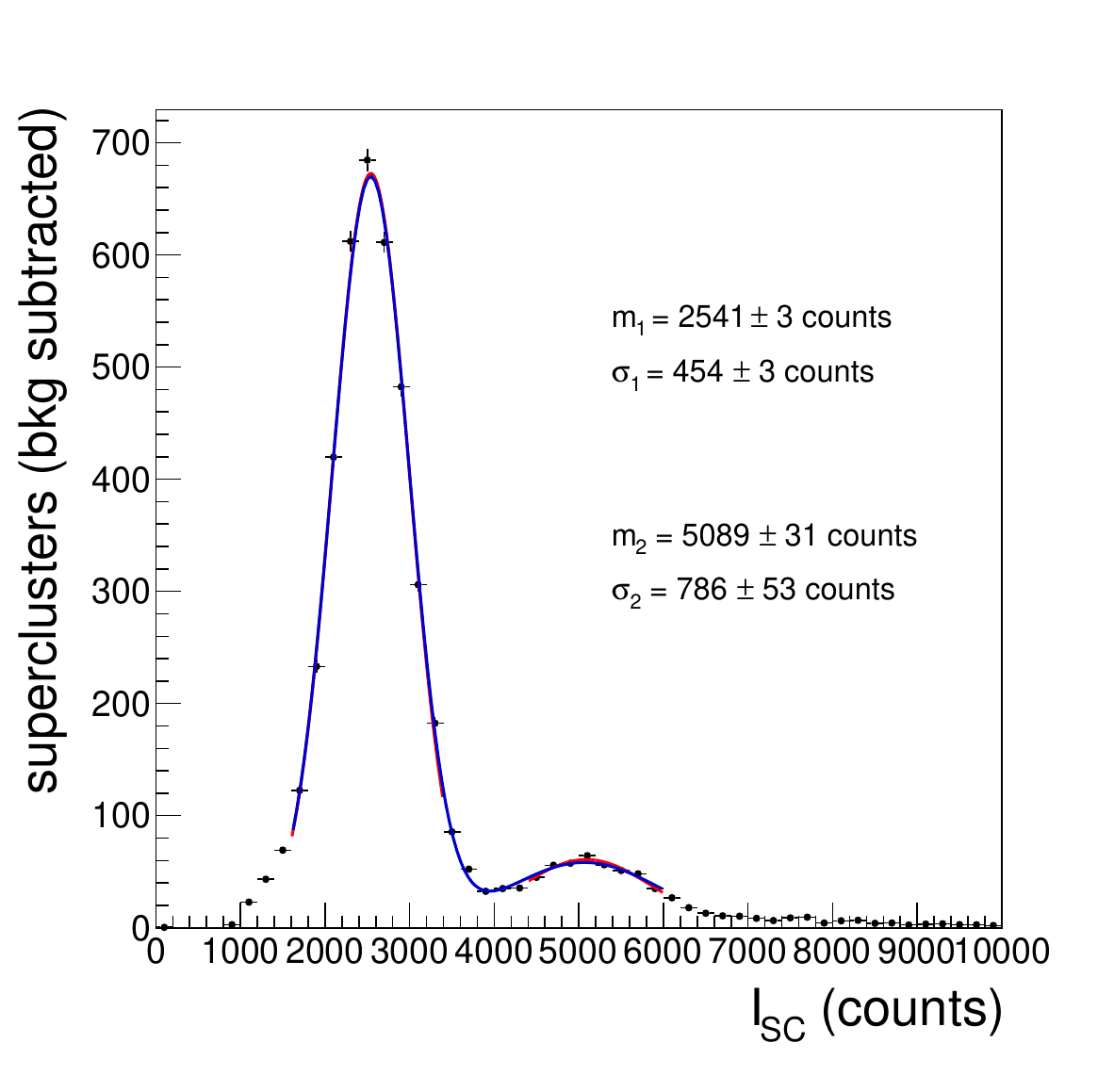}
    \caption{Distribution of the supercluster integral, before the
      absolute energy scale calibration is applied, in events with the
      \fe source. Clearly visible is the large peak of a single spot,
      and, at around twice the energy, a broader peak for the case of
      two neighboring spots merged in a single supercluster.
      \label{fig:feuncalibpeak}}
  \end{center}
\end{figure}

The absolute energy scale was then calibrated with the energy
distribution measured in data with the \fe source, which provides
monochromatic photons of 5.9\keV, with the procedure described in
Ref.~\cite{bib:fe55}. The supercluster integral is defined as:
\begin{equation}
\label{eq:integral}
I_{SC} = \sum_i^{cluster} N_{ph}^i,
\end{equation}
where $N_{ph}^i$ is the number of counts (photons) in the $i^{th}$
pixel, and the sum runs over all the pixels of the supercluster.
While to perform the basic- and super-clustering only pixels passing
the zero suppression are considered, for the energy estimate in
Eq.~\ref{eq:integral} all the pixels within the cluster contours are
counted, eventually having negative $N_{ph}^i$ after the pedestal
subtraction. This choice is meant to avoid a bias on the energy
estimate, since after the pedestal subtraction the distribution of the
noise is centered around zero.  The distribution of \isclu, for a run
taken in presence of \fe source, is shown in
Fig.~\ref{fig:feuncalibpeak}. In addition to the main peak, with a
mean of about 2500\unit{counts}, a broader peak is clearly
distinguished, which represents the cases of two merged spots, with an
integral twice the single spot one. An example of such a merged spot
is given in Fig.~\ref{fig:basic_clusters} (left).
 The position of the maximum in the single-spot distribution in runs
with \fe source allowed to calibrate the absolute energy scale of
the \lemon detector.  The energy resolution for the reconstructed \gac
superclusters is about 18\%, similar to the one that can be obtained
with only the basic clustering step with \idbscan~\cite{iDBSCAN}, and
improving the one with the simple \nnc algorithm previously
used~\cite{bib:fe55}.

Using runs with this monochromatic, high rate source, positioned at
different distances from the GEM planes, a decrease of the light
response for lower distances from the GEM was observed. This effect is
opposite to the expected behavior of a lower light yield at larger
distances. Indeed, it is expected that, during the drift along the
$z$-direction, the ionization charge undergoes a diffusion in the TPC
gas, and some electrons are removed by attachment to the gas molecule.
Consequently, some loss in the light collection may be expected. The
opposite behavior, instead, is clearly observed. While this effect is
currently under study in more detail, it was attributed to a possible
saturation effect of the GEMs, especially in the third stage of
multiplication, where the charge density in one GEM hole is
maximal.  Under this hypothesis, an effective, empirical correction
was developed, which relies on the charge density of a cluster from
a \fe deposit. The light density, $\delta$, is defined as:
\begin{equation}
  \label{eq:density}
  \delta = \isclu / n_p,
\end{equation}
where $n_p$ is the number of pixels passing the zero-suppression
threshold (differently from the definition of \isclu, where all the
pixels in the supercluster are considered). This effective calibration
returns the absolute energy of a spot-like region, similar in size to
the \fe clusters, as a function of the supercluster density, $\delta$:
$E=c(\delta)\cdot I_{SC}$. In the hypothesis of saturation, the
\textit{local} density along the track is the parameter which
regulates the magnitude of the effect, thus the correction has to be
applied dynamically for slices of the supercluster having a size
similar to the \fe spots.  This is achieved with the procedure
described in the following.

\begin{figure}[ht]
  \begin{center}
     \includegraphics[width=0.49\linewidth]{figures/pic_run02317_ev8_oriIma_paper_zoom}
     \includegraphics[width=0.49\linewidth]{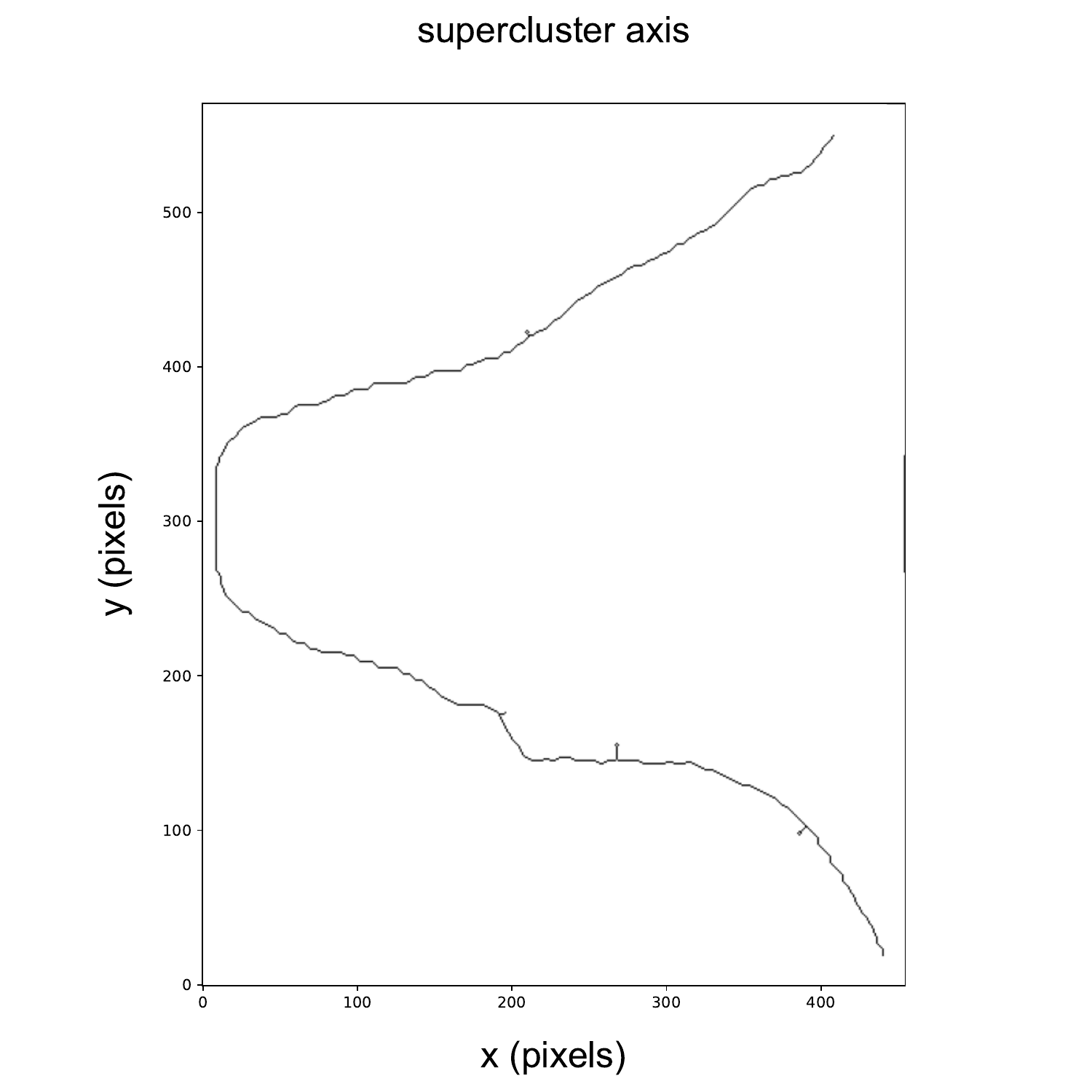}

      \caption{Left: zoom on the full-resolution image of a track
        candidate in a run with the \ambe radioactive source. Right:
        output of the skeletonization and pruning of the branches for
        one example supercluster extended in
        space.  \label{fig:skeleton}}

    \end{center}
\end{figure}

First, the supercluster \textit{skeleton}, \ie, the 1-pixel-wide
representation along the path, is reconstructed.  This is achieved
through a morphological thinning of the superclusters with the iterative
algorithm from Ref.~\cite{thin1,thin2}.  Second, a pruning of the
obtained skeleton is done, to remove residual small branches along the
main pattern, using a hit-or-miss transform. The output of this
process for one example track is shown in Fig.~\ref{fig:skeleton}.
For the calibration procedure, the found skeleton is followed,
starting from one of the two end points, and circles having their
center on a pixel of the skeleton and their radius equal to the
average spot size of the \fe clusters are defined. It was checked
that this procedure includes all the pixels of the original cluster
for the vast majority of the clusters considered.  The local density
$\delta_s$ of the slice $s$ is computed, and its integral $I_s$ is
calibrated to an absolute energy through the effective correction
$E_s=c(\delta_s)\cdot I_s$. The pixels of the supercluster used for
the slice calibration are removed (including the skeleton ones), and
the procedure is iterated, until having included all the pixels. The
sum of the energies of all the slices is the estimate of the
calibrated energy of the supercluster:
\begin{equation}
  \label{eq:ecal}
  E_{SC} = \sum_s^{slices} E_s
\end{equation}
As a closure test of this procedure, the calibrated energy of the
superclusters reconstructed in the runs with the \fe source is
obtained.  The value of the energy peak was obtained by fitting the
distribution with the same function used in
Fig.~\ref{fig:feuncalibpeak}, and equals to  $m_1 = 5.93 \pm 0.01$\keV,
compatible with the expected value. The calibration procedure is an
overkill for the case of the small \fe spots, but it is necessary for
very long cosmic ray tracks or even for medium-length superclusters
from nuclear and electron recoils.  The energy resolution worsen after
the calibration ($\sigma_1 = 1.48 \pm 0.01$, \ie, 25\% energy
resolution), as a sign that the empiric correction is still
suboptimal.

The skeletonization procedure provides a general method to estimate
the track length ($l_p$), accurate both in the case of straight and
curving track. As a check, it has been verified that, in the case of
straight tracks, the length extracted in this way coincides with the
length of the major axis estimated with a singular value decomposition
(SVD), described in the following section. For exactly round spots,
the skeleton would collapse in the center of the cluster and the
resulting length would be 1 pixel, but this completely symmetric case
never happens in the considered samples.

\section{Cluster shape observables}
\label{sec:clustershapes}
The interaction of different particles with the nuclei or the
electrons in the gas of the TPC produces different patterns of the 2D
projection of the initial 3D particle trajectory.  These cluster shape
observables are useful to discriminate different ionizing
particles. In particular, they were used to select a pure sample of
nuclear recoil candidates produced by the interaction of the neutrons
originating from the \ambe source and to identify various sources of
backgrounds. The main cluster shape observables are described in the
following:

\begin{itemize}
  \item \textit{projected length and width:~} a SVD on the $x \times
    y$ matrix of the pixels belonging to the supercluster is
    performed. The eigenvectors found can be interpreted as the
    directions of the two axes of an ellipse in 2D. The eigenvalues
    represent the magnitudes of its semiaxes: the major one is defined
    as \textit{length}, $l$ the minor one as \textit{width},
    $w$. These are well defined for elliptic clusters, or for long and
    straight tracks. The directions along the major and the minor axis
    are defined as \textit{longitudinal} and \textit{transverse} in
    the following. The longitudinal and transverse supercluster
    profiles, for the cosmic ray track candidates shown as an example
    in Fig.~\ref{fig:super_clusters2} (bottom) are shown in
    Fig.~\ref{fig:profiles}. The longitudinal profile shows the
    typical pattern of energy depositions in clusters, while the
    transverse profile, dominated by the diffusion in the gas, shows a
    Gaussian shape. It has to be noted that the cluster sizes
    represent only the projection of the 3D track in the TPC on the 2D
    $x$--$y$ plane;

    \begin{figure}[ht]
      \begin{center}
        \includegraphics[width=0.45\linewidth]{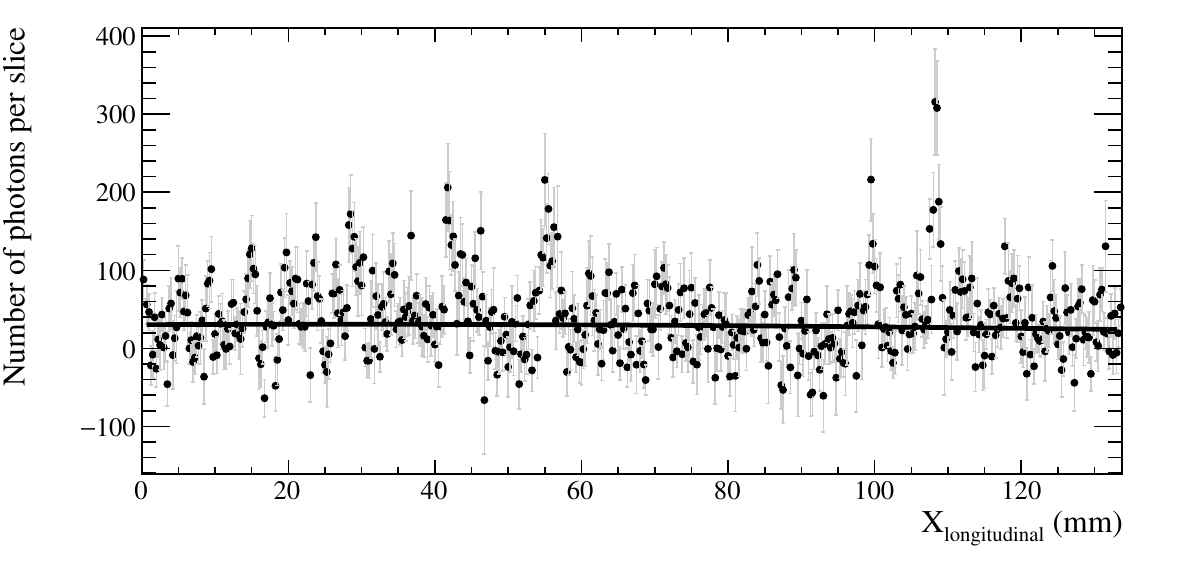}
        \includegraphics[width=0.45\linewidth]{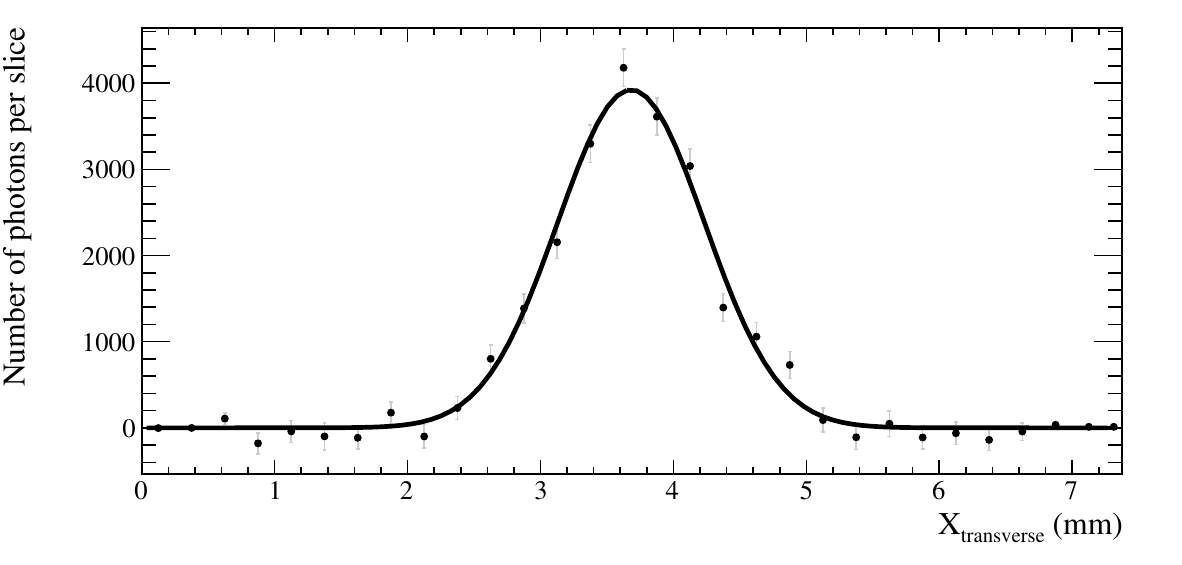}
        \caption{Supercluster profile in the longitudinal (left) or
          transverse (right) direction, for a long and straight cosmic
          ray track candidate shown in Fig.~\ref{fig:super_clusters2}
          (bottom). The longitudinal profile shows an energy
          deposition in sub-clusters, while the transverse direction
          shows the typical width of the diffusion in the gas. For the
          longitudinal profile, the line represent the average number
          of photons per slice. For the transverse profile, it
          represents a fit with a Gaussian
          PDF. \label{fig:profiles}}
      \end{center}
  \end{figure}

  \item \textit{projected path length:~} for curly and kinky tracks
    the values returned by the SVD of the supercluster are not an
    accurate estimates of their size. While the width is dominated by
    the diffusion, the length for patterns like the one shown in the
    example of Fig.~\ref{fig:super_clusters1} is ill-defined. Thus,
    the more general path length, $l_p$, computed with the
    skeletonization procedure in Fig.~\ref{fig:skeleton} is used to
    estimate the linear extent for both straight and curved tracks.

  \item \textit{Gaussian width:~} the original width of the track in
    the transverse direction is expected to be much lower than the
    observed width induced by the diffusion in the gas. Thus, as shown
    in Fig.~\ref{fig:profiles} (right), the standard
    deviation, \tsigmag, can estimated by a fit with a Gaussian
    probability density function (PDF). This modelling is appropriate
    for the cases of straight tracks, typical of cosmic-ray
    background, or energetic nuclear recoils, and also for spot-like
    clusters. Tracks with kinks are badly modeled by a Gaussian
    function: if the $\chi^2$ of the convergence of the fit is below a
    threshold, the estimated width from the SVD method described
    earlier is used;

  \item \textit{slimness:~} the ratio of the width over the path
    length, $\xi=w/l_p$, represents the aspect ratio of the
    cluster. It is very useful to discriminate between
    cosmic-ray-induced background (long and thin) from low energy
    nuclear or electron recoils (more elliptical or round, as the \fe
    spots);
    
  \item \textit{integral:~} the total number of photons detected by all the
  pixels gathered in the supercluster, \isclu, as defined in
  Eq.~\ref{eq:integral};

  \item \textit{pixels over threshold:~} the number of pixels in the
  supercluster passing the zero-suppression threshold, $n_p$;

  \item \textit{density:~} the ratio $\delta$ of \isclu, divided by
  $n_p$, as defined in Eq.~\ref{eq:density};

  \item \textit{energy:~} the calibrated energy, expressed in keV. The
    calibration method simultaneously performs both the per-slice
    correction as a function of the local $\delta$, and the absolute
    energy scale calibration, which corrects the non perfect
    containment of the cluster, \ie, the bias in the distribution of
    $E/E_{true}$, using with \fe source.
\end{itemize}

\begin{figure}[ht]
  \begin{center}
  \includegraphics[width=0.45\linewidth]{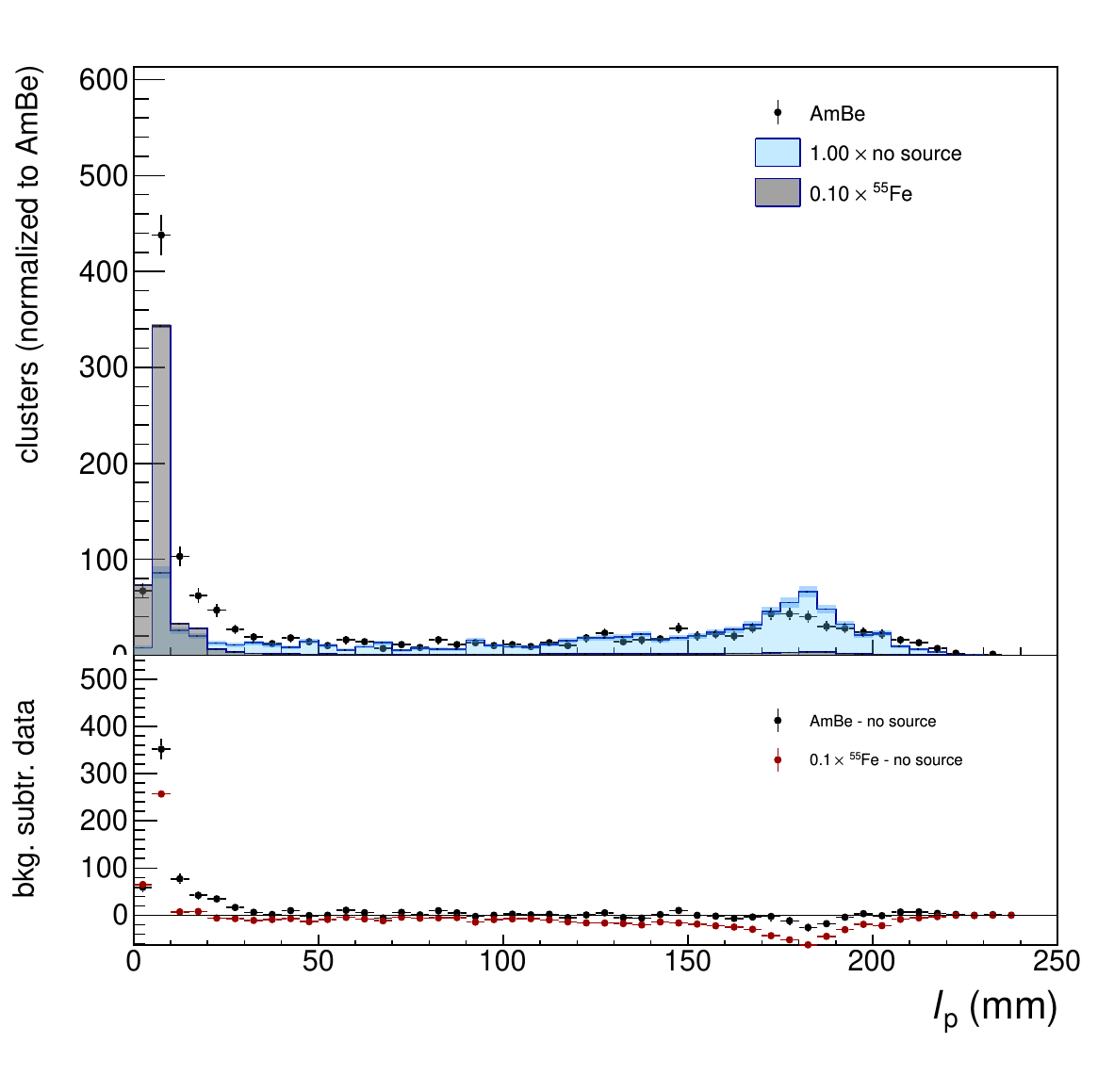}
  \includegraphics[width=0.45\linewidth]{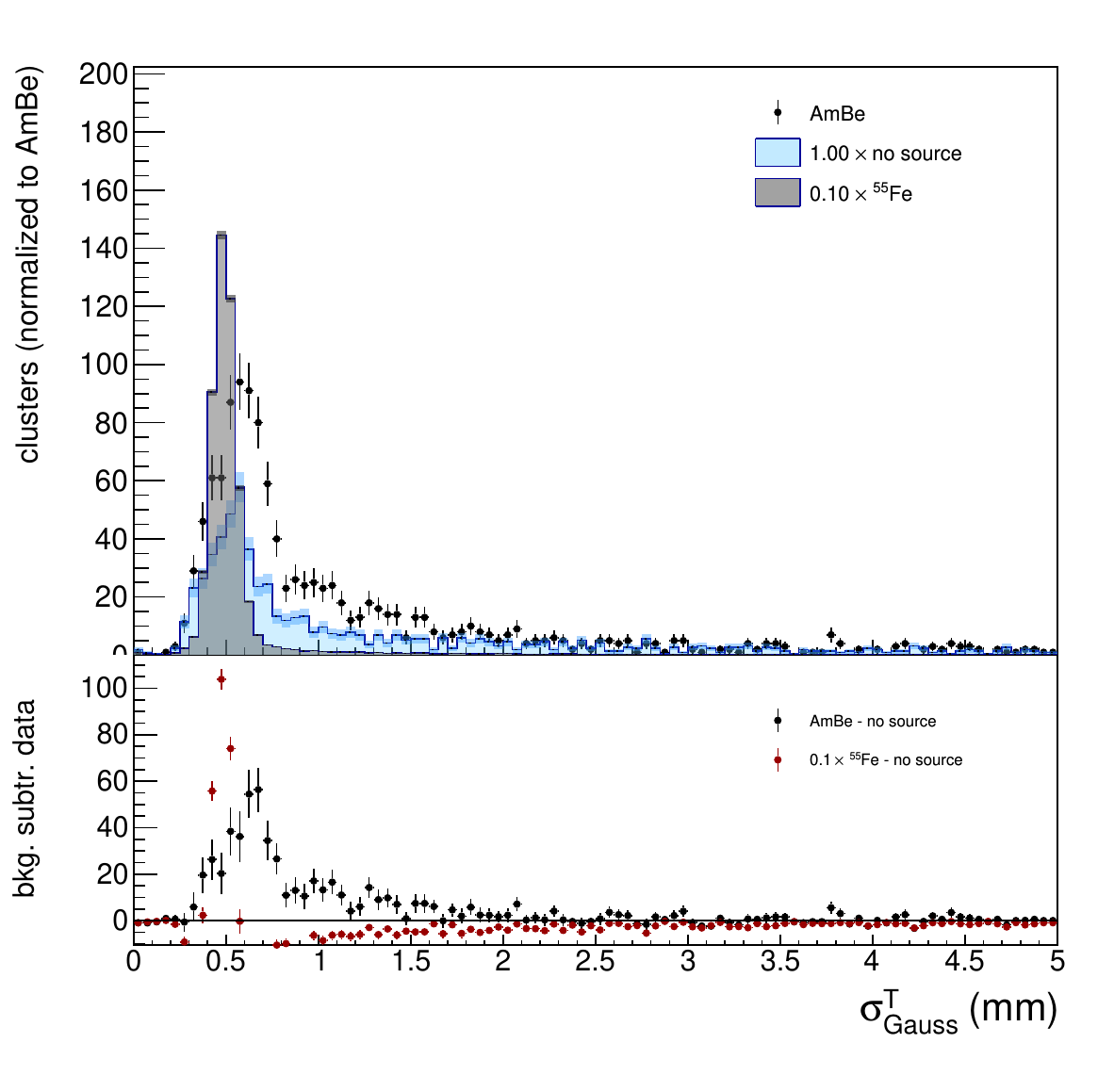}

  \caption{Supercluster sizes projected onto the $x$--$y$ plane. Left:
    longitudinal path length, $l_p$.  Right: transverse Gaussian
    spread, \tsigmag. Filled points represent data with \ambe source,
    dark gray (light blue) distribution represents data with \fe
    source (no source).  The normalization of data without any
    radioactive source is scaled to the same exposure time of
    the \ambe one. For the data with \fe source , a scaling factor of
    one tenth is applied for clearness, given the larger activity of
    this source. Filled dark-grey and dark-azure bands represent the
    statistical uncertainty on data with no source and \fe source,
    respectively. \label{fig:clsize}}

    \end{center}
\end{figure}

The distributions of the projected supercluster path length, $l_p$,
and Gaussian transverse size, \tsigmag, are shown in
Fig.~\ref{fig:clsize}, for data taken in different types of runs.
During the data-taking approximately 3000 frames were recorded in
absence of any external radioactive source ({\it no-source}
sample). In these frames the interaction of ultra-relativistic cosmic
ray particles (mostly muons) are clearly visible as very long
clusters. Internal radioactivity of the \lemon materials also
contributes with several smaller size clusters. About 1500 frames were
acquired with the \ambe source, and approximately $10^4$ calibration
images with \fe source. In Fig.~\ref{fig:clsize}, as well as in the
following ones showing other cluster properties, the distributions
obtained in runs without radioactive sources are normalized to
the \ambe data total CMOS exposure time. For the data with \fe source,
since the activity of the source is such to produce about 15
clusters/event, the data are scaled by a factor one-tenth with respect
to the \ambe exposure time for clearness. Tracks originating from
cosmic-ray interactions are present also in the data with both types
of radioactive sources, with an efficiency scale factor slightly
different from unity because of the trigger. Considerations about the
trigger efficiency, affecting the background normalization in data
with a radioactive source are given in Sec.~\ref{sec:background}. The
distributions in this section aim to show the different cluster shape
observables among the different kinds of events.

Figure~\ref{fig:clsize} shows the cluster sizes distributions in the
longitudinal and transverse directions for different sets of
runs. Data show an average Gaussian width for the \fe spots
$\tsigmag\approx500$\unit{$\mu$m} (dominated by the diffusion in the
gas), while it is larger, approximately 625\unit{$\mu$m}, for data
with \ambe source.  The contribution of cosmic rays, present in all
the data, is clearly visible in the data without any radioactive
source, corresponding to clusters with a length similar to the
detector transverse size (22\unit{cm}).

\begin{figure}[ht]
  \begin{center}
  \includegraphics[width=0.45\linewidth]{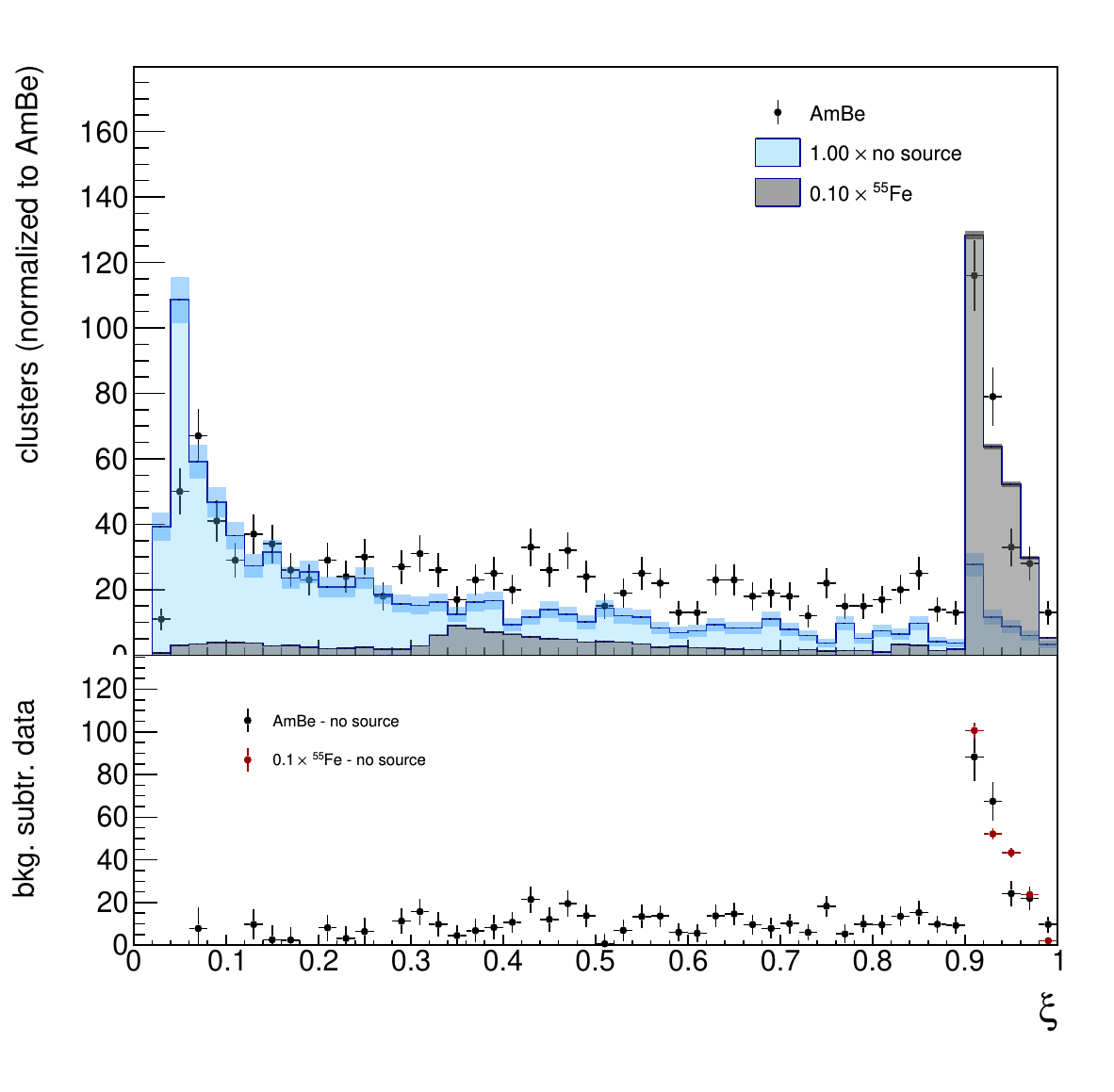}
  \includegraphics[width=0.45\linewidth]{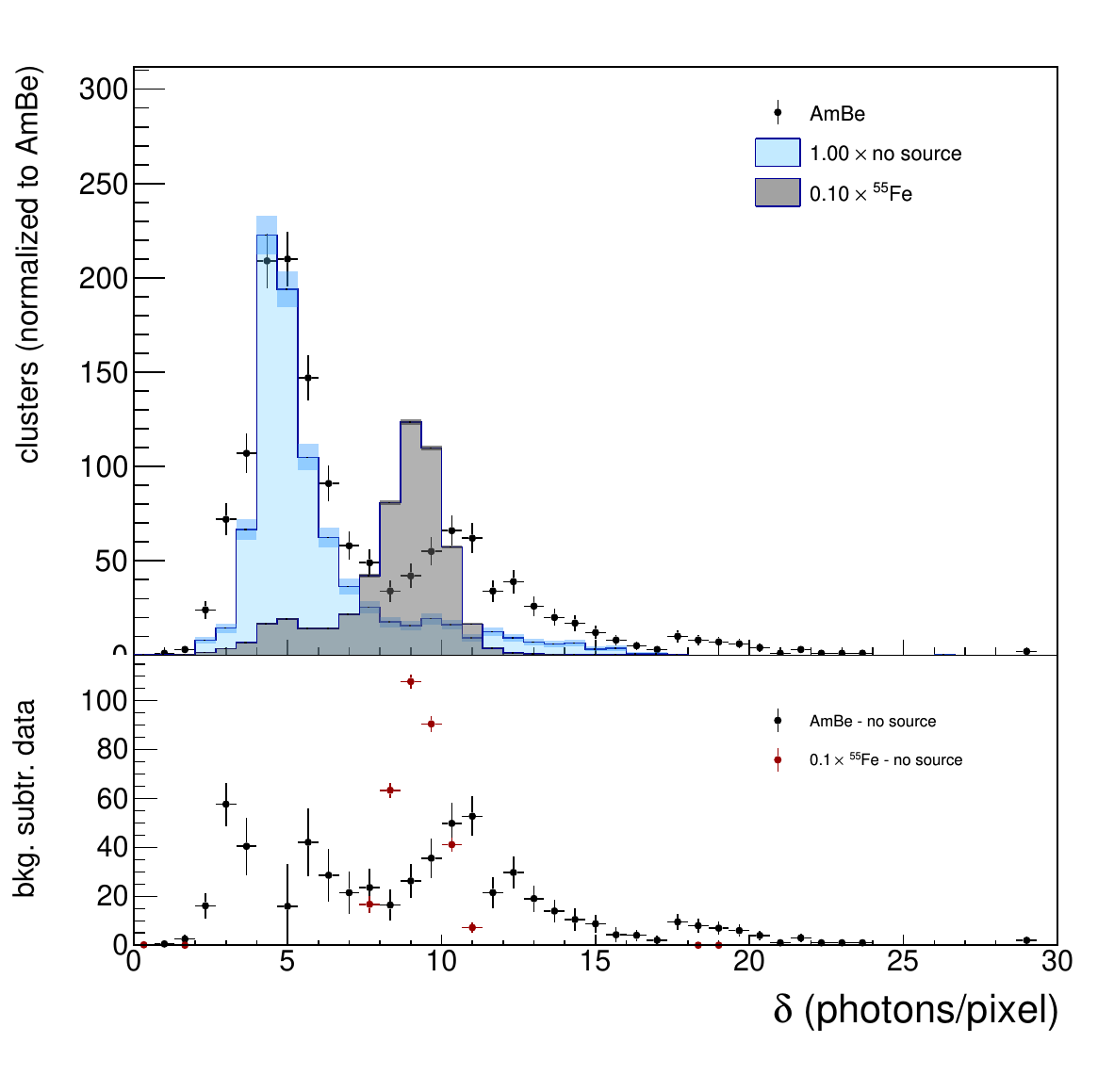}

  \caption{Supercluster variables. Left: slimness $\xi$; right: light
    density $\delta$. Filled points represent data with \ambe source,
    dark gray (light blue) distribution represents data with \fe
    source (no source).  The normalization of data without source is
    to the same exposure time of the \ambe one. For the data with \fe,
    a scaling factor of one tenth is applied for clearness, given the
    larger activity of this source. Filled dark-grey and dark-azure
    bands represent the statistical uncertainty on data with no source
    and \fe source, respectively. \label{fig:clshape}}

\end{center}
\end{figure}

Other observables are the slimness, $\xi$, and the light density,
$\delta$, shown in Fig.~\ref{fig:clshape}. The former is a useful
handle to reject tracks from cosmic rays, which typically have a slim
aspect ratio, \ie, low values of $\xi$, while the clusters from \fe
are almost round, with values $0.9\lesssim\xi<1$. By construction,
$\xi<1$, since the width is computed along the minor axis of the
cluster, and for round spots it peaks at around 0.9. The apparent
threshold effect is purely geometrical, due to the minimal size of the
macro-pixel ($4{\times}4$) used at the basic clustering step which can
be larger than a round spot from \fe. Data with \ambe source, which
contains a component of nuclear recoils, show a component of round
spots, similar in size to the ones of \fe, and a more elliptical
component, with $0.4<\xi<0.8$ values. Aside from the slightly
different normalization of the cosmic-ray background in the region
dominated by these interactions, $0<\xi<0.3$, which is discussed in
Sec.~\ref{sec:background}, the shape of the distribution shows some
differences. This can be attributed to the cases of clusters induced
by the radiation from the source which may overlap with a long track
from a cosmic ray, changing its shape or, rarely, inducing a track
splitting.  The effect on the following statistical analysis is an
underestimation of the signal efficiency, since a loose selection on
this variable is used to discard cosmic-rays background. Given that
the signal concentrates at values of $\xi \approx 0.9$, this effect is
assumed to be negligible.

Finally, the light density, $\delta$, is the variable expected to
better discriminate among different candidates: cosmic-ray-induced
background, electron recoils and nuclear recoil candidates. This is
the variable used in this paper for the final particle identification.
Due to electron diffusion in gas along the drift path, the measured
$\delta$ depends on the $z$ position of the cluster.

Because the spot size is expected to increase by about a factor 4 from 
$z$~=0\unit{cm} to $z$~=20\unit{cm} (equation~\ref{eq:diff}), 
assuming an uniform distribution of clusters within the sensitive volume, 
size distribution is expected to have a relative standard deviation of about 35\%.
This spoils the precision in the evaluation of the real $\delta$ 
both for the signal and for the background cluster resulting in a deterioration of the selection performance.
Different methods to evaluate the cluster $z$ are under development and will
allow a more effective determination of $\delta$.

Moreover, the identification results can be improved using
additional cluster shape variables, also profiting of their different
correlations for signal and background clusters, via a multivariate
approach, but in this early analysis a simple cut-based approach is
chosen, for simplicity.

\begin{figure}[ht]
  \begin{center}
  \includegraphics[width=0.45\linewidth]{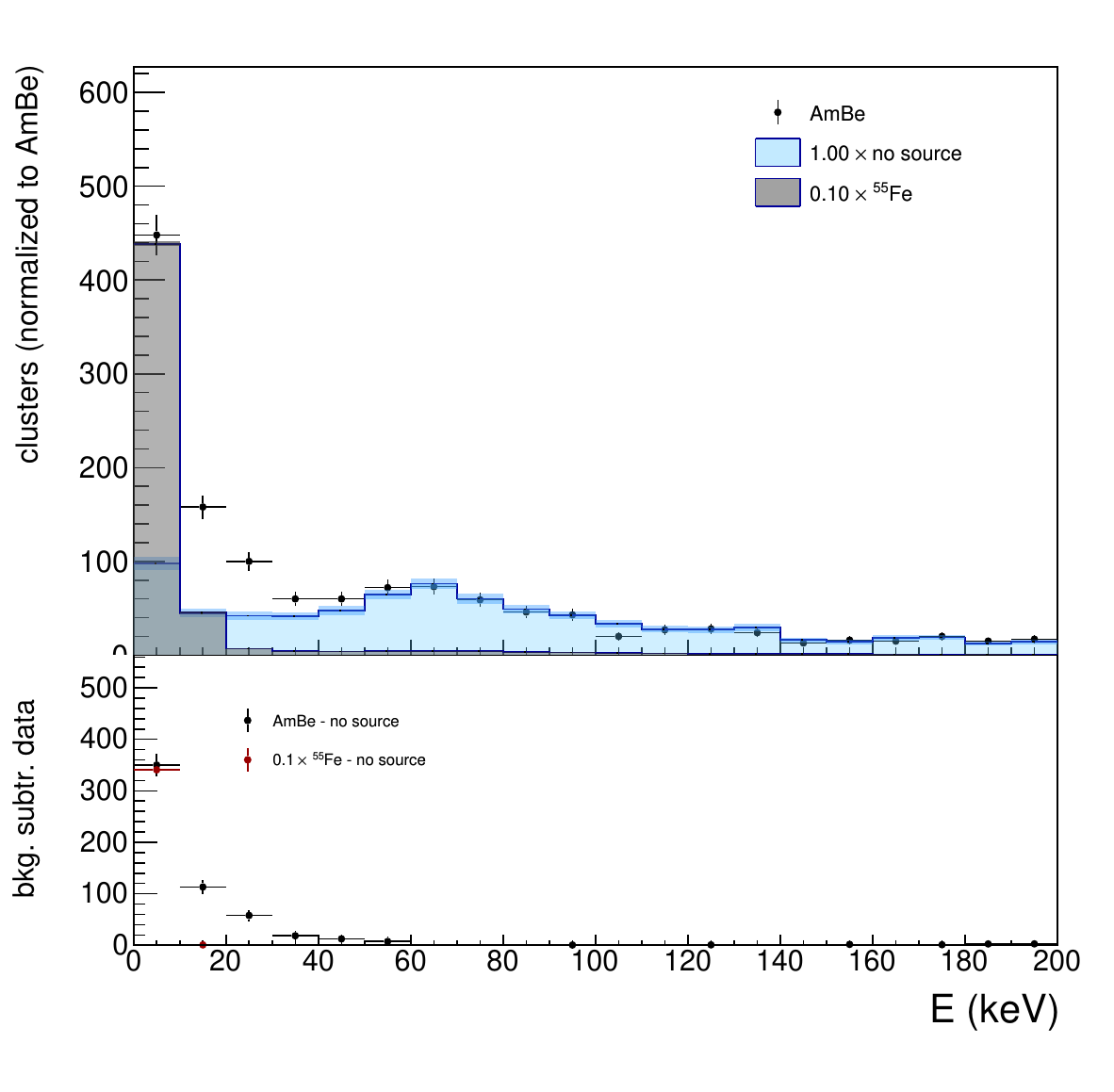}
  \includegraphics[width=0.45\linewidth]{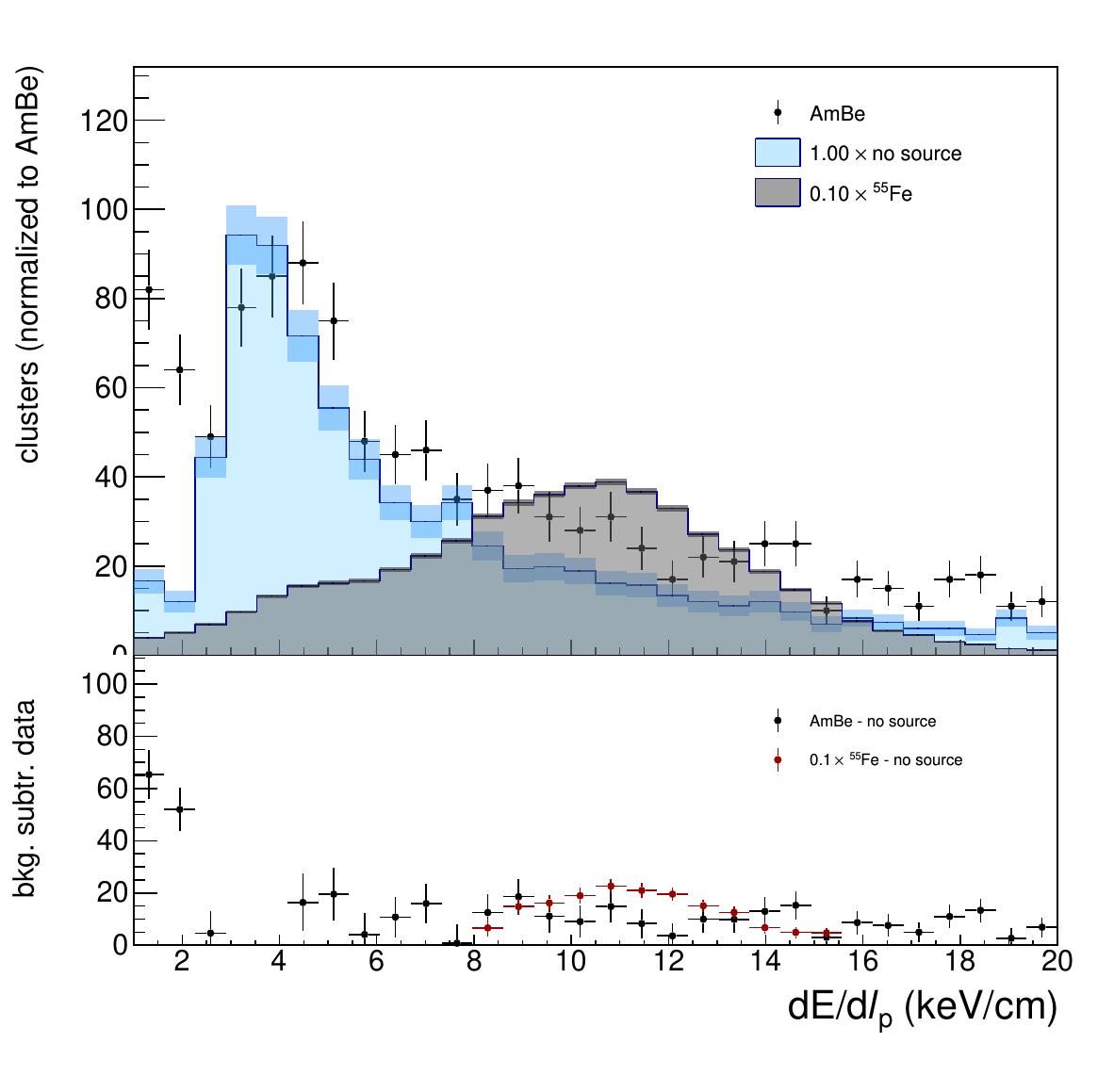}

  \caption{Supercluster calibrated energy spectrum (left) and their
    average \dedl. Filled points represent data with \ambe source,
    dark gray (light blue) distribution represents data with \fe
    source (no source).  The normalization of data without source is
    to the same exposure time of the \ambe one. For the data with \fe,
    a scaling factor of one tenth is applied for clearness, given the
    larger activity of this source. Filled dark-grey and dark-azure
    bands represent the statistical uncertainty on data with no source
    and \fe source, respectively. \label{fig:energy}}

\end{center}
\end{figure}

Finally, Fig.~\ref{fig:energy} (left) shows the calibrated energy
($E$) spectrum for the reconstructed superclusters. The energy
spectrum shows the $E=5.9\keV$ peak in the first bin of the
distribution for data with \fe source, and the expected broad peak for
minimum ionizing particles traversing the $\approx$20\unit{cm} gas
volume at around 60\keV. The distribution of the observed average
projected \dedl for the no-source sample and for the \ambe samples is
shown in Fig.~\ref{fig:energy} (right). The broadening of the
distribution is mainly due to the specific energy loss fluctuation in
the gas mixture of the cosmic ray particles.  Its modal value,
corrected for the effect of the angular distribution (an average
inclination of 56$^{\circ}$ was measured from track reconstruction) is
2.5\keV/cm, in good agreement with the \garfield prediction of
2.3\keV/cm.

\subsection{Background normalization}
\label{sec:background}
The data with \ambe source, taken on the Earth surface, suffers from a
large contribution of interactions of cosmic rays, and from ambient
radioactivity, whose suppression is not optimized for the \lemon
detector. The cluster shape observables provide a powerful handle to
discriminate them from nuclear recoils candidates, but the small
residual background needs to be statistically subtracted. The
distributions shown earlier, where the different types of data are
normalized to the same exposure time, demonstrate that the live-time
normalization provides already a good estimate of the amount of cosmic
rays in data with radioactive sources. This approach does not account
for a possible bias from the trigger, which is generated by the PMT
signals, as described in Sec.~\ref{sec:layout}. Indeed, in runs with
the \ambe source, the PMT can trigger both on signals from neutron
recoils or photons produced by the $^{241}$Am, and on ubiquitous
signals from cosmic rays, while in the sample without source only the
latter are possible.  Therefore, during the same exposure time,
the probability to trigger on cosmic rays is lower in events
with \ambe than in no-source events. The trigger efficiency scale
factor, $\varepsilon_{SF}$, can be obtained as the ratio of the number
of clusters selected in pure control samples of cosmic rays ($CR$)
obtained on both types of runs:
\begin{equation}
\label{eq:sfeff}
\varepsilon_{SF} = \frac{N^{AmBe}_{CR}}{N^{no-source}_{CR}}.
\end{equation}

The $CR$ control region is defined by selecting clusters with
$l>13$\unit{cm}, $\xi<0.1$, $\tsigmag<6$\unit{mm}, and having an
energy within a range dominated by the cosmic rays contribution,
$50<E<80\keV$. The selected clusters show small values of
$\delta\approx5$, well compatible with the small specific ionization
of ultra-relativistic particles.  This sample is limited in
statistics, but it is expected to be almost 100\% pure. The scale
factor obtained is $\varepsilon_{SF}=0.75\pm0.02$.  A possible
consequence of occasional nearby clusters from nuclear recoils
changing the distribution of $\xi$ is considered. The effect is
expected to be reduced for the long clusters selected in the $CR$
control region. A systematic uncertainty due to the different shape of
$\xi$ is estimated by applying the $CR$ selection on the \ambe
clusters where the slimness is substituted, cluster-by-cluster,
sampling the distribution in the no-source data. The difference in the
$N^{AmBe}_{CR}$ is about 1\%, small with respect to the statistical
uncertainty of $\varepsilon_{SF}$.

In addition to the normalization correction of the cosmic-ray
background to be applied before subtracting it from the source data,
it has been mentioned earlier that occasional overlap of long tracks
from this background and signal tracks moderately distort the cluster
shape distributions. Splitting of a long track in the data with the
source may happen because of a gradient change in the region of the
overlap with a dense nuclear recoil cluster. The typical track lengths
of cosmic rays and signal tracks are so different that in these
occurrences the split tracks are still much longer than the signal
ones, thus they are rejected.  In other cases, when the overlap is
close to the edges, the track length is increased, and the selection
dismisses those clusters.  Since the requirement on the track length
is very loose, the effect on the signal efficiency is assumed to be
negligible.

\begin{figure}[ht]
  \begin{center}
  \includegraphics[width=0.45\linewidth]{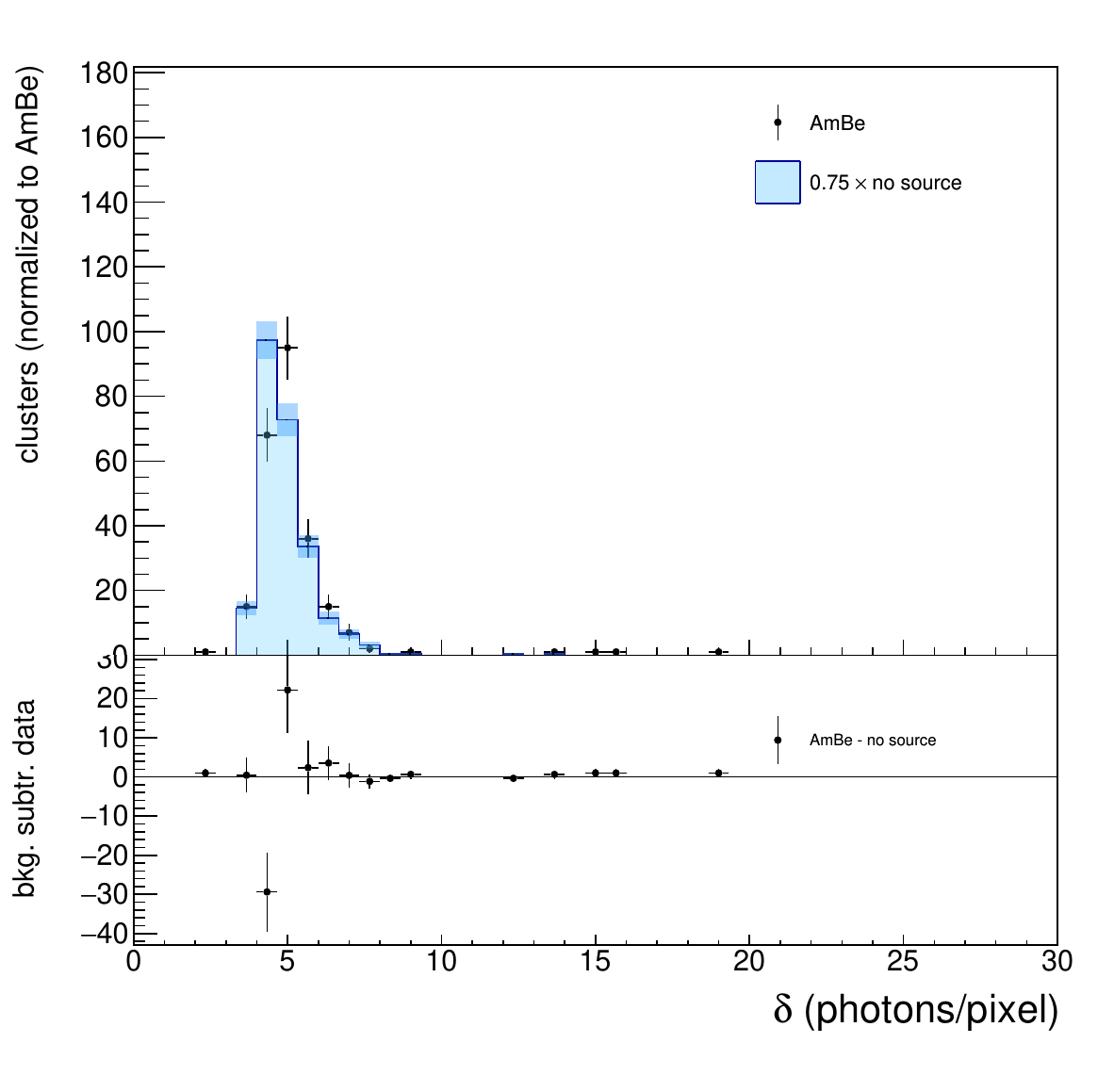}
  \includegraphics[width=0.45\linewidth]{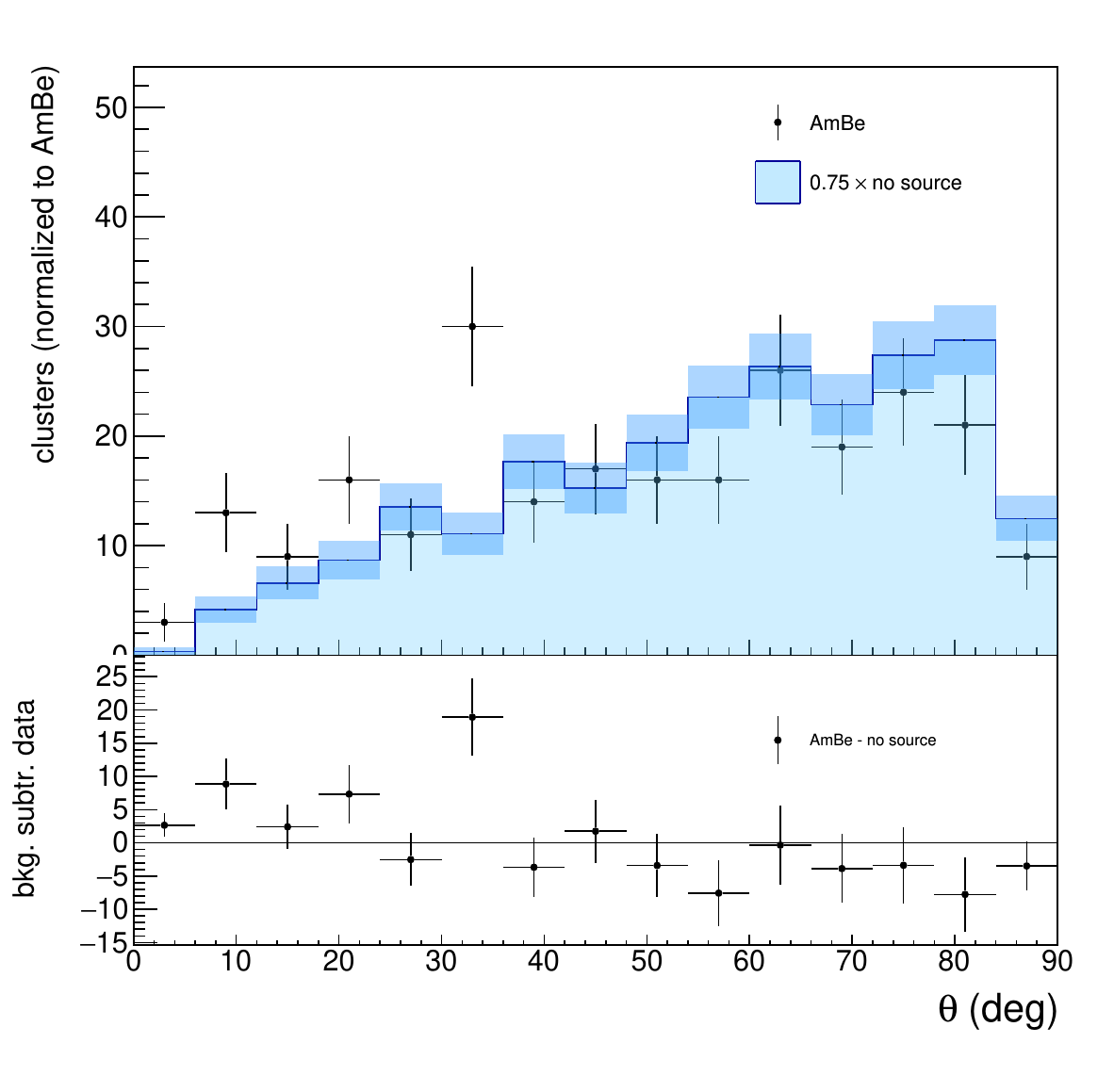}

  \caption{Supercluster light density $\delta$ (left) and polar angle
    (right) - with respect the horizontal axis - distributions for
    long clusters, dominated by cosmic rays tracks. The clusters are
    selected by the $CR$ requirements, except the energy one.  Filled
    points represent data with \ambe source, light blue distribution
    represents data without any radioactive source.  The normalization
    of data without source is to the same exposure time of the \ambe
    one, accounting for the trigger scale factor $\varepsilon_{SF}$,
    as defined in the text. Bins with negative contents are set to
    zero value. The filled dark-azure band represents the statistical
    uncertainty on data with no source. \label{fig:cosmics}}

\end{center}
\end{figure}

In Fig.~\ref{fig:cosmics} the typical light density and polar angle
(with respect the horizontal axis) distributions for long clusters
selected with the $CR$ requirements, except the energy one are shown
for the \ambe and for the no-source sample, after having applied the
$\varepsilon_{SF}$ scale factor to the latter.  Clusters with
$\delta<6$ are thus expected to be mostly coming from muon tracks, and
they show indeed a polar angle which is shifted at values towards
$90^\circ$.

\section{Nuclear recoil identification results}
\label{sec:results}
As mentioned in the previous Section, the 1D observable chosen to
distinguish the signal of nuclear recoils from the various types of
background is the energy density $\delta$ of the cluster.

\subsection{Signal preselection}
To enhance the purity of the signal sample, a preselection was
applied, prior to a tighter selection on $\delta$: clusters with
$l_p>6.3$\unit{cm} or $\xi<0.3$ were rejected to primarily suppress
the contribution from cosmic rays. A further loose requirement
$\delta>5$ photons/pixel was also applied to remove the residual
cosmic rays background based on their low specific ionization. These
thresholds, which only reject very long and narrow clusters, are very
loose for nuclear recoils with $E<1\MeV$ energies, given the expected
range in simulated events, shown in Fig.~\ref{fig:range}, of less than
1\unit{cm}. Thus the preselection efficiency for signal is assumed to
be 100\%. For electron recoils it can be estimated on data by using
the \fe data sample, and is measured to be
$\varepsilon_{B}^{presel}=70\%$. Since the X-ray photo-electrons of
this source are monochromatic, the estimate of the electron recoils
rejection is only checked for an energy around $E=5.9$\keV. The
spectrum of nuclear recoils from \ambe source, instead, extends over a
wider range of energies, around [1--100]\keV.

\begin{figure}[ht]
  \begin{center}
  \includegraphics[width=0.90\linewidth]{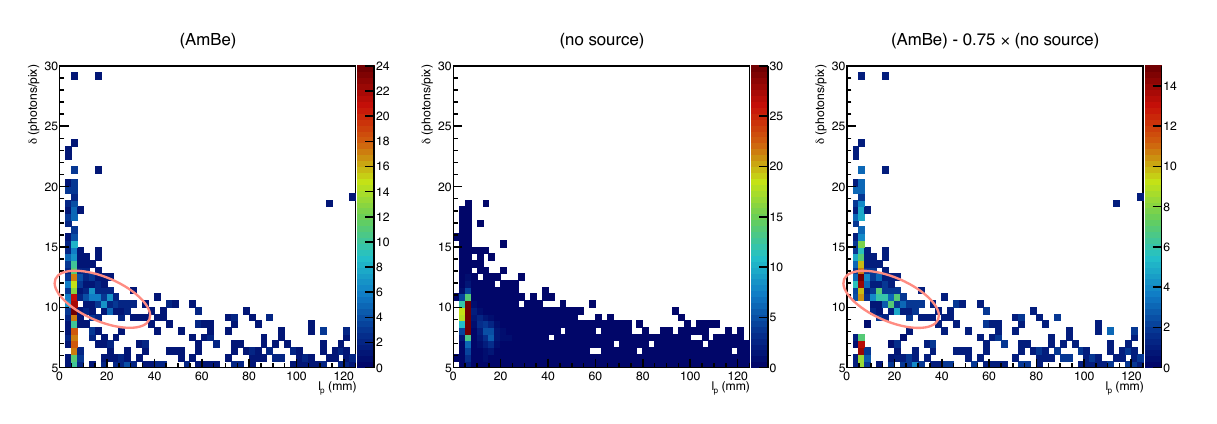}

  \caption{Supercluster light density $\delta$ versus length $l_p$,
    for data with \ambe source (left), data without any artificial
    source (middle), and the resulting background-subtracted \ambe
    data.  The normalization of data without source is to the same
    exposure time of the \ambe one, accounting for the trigger scale
    factor $\varepsilon_{SF}$, as defined in the text. The orange
    ellipse represents the approximate contour of the 59\keV photons
    control region ($PR$) defined in the text. \label{fig:dvsl}}

  \end{center}
\end{figure}

With this preselection, the distribution in the 2D plane
$\delta$--$l_p$ is shown in Fig.~\ref{fig:dvsl} for \ambe source and
no-source data and for the resulting background-subtracted \ambe data.
The latter distribution shows a clear component of clusters with short
length ($l_p\lesssim1$\unit{cm}) and high density ($\delta\gtrsim
10$), expected from nuclear recoils deposits.

\begin{figure}[ht]
  \begin{center}
  \includegraphics[width=0.60\linewidth]{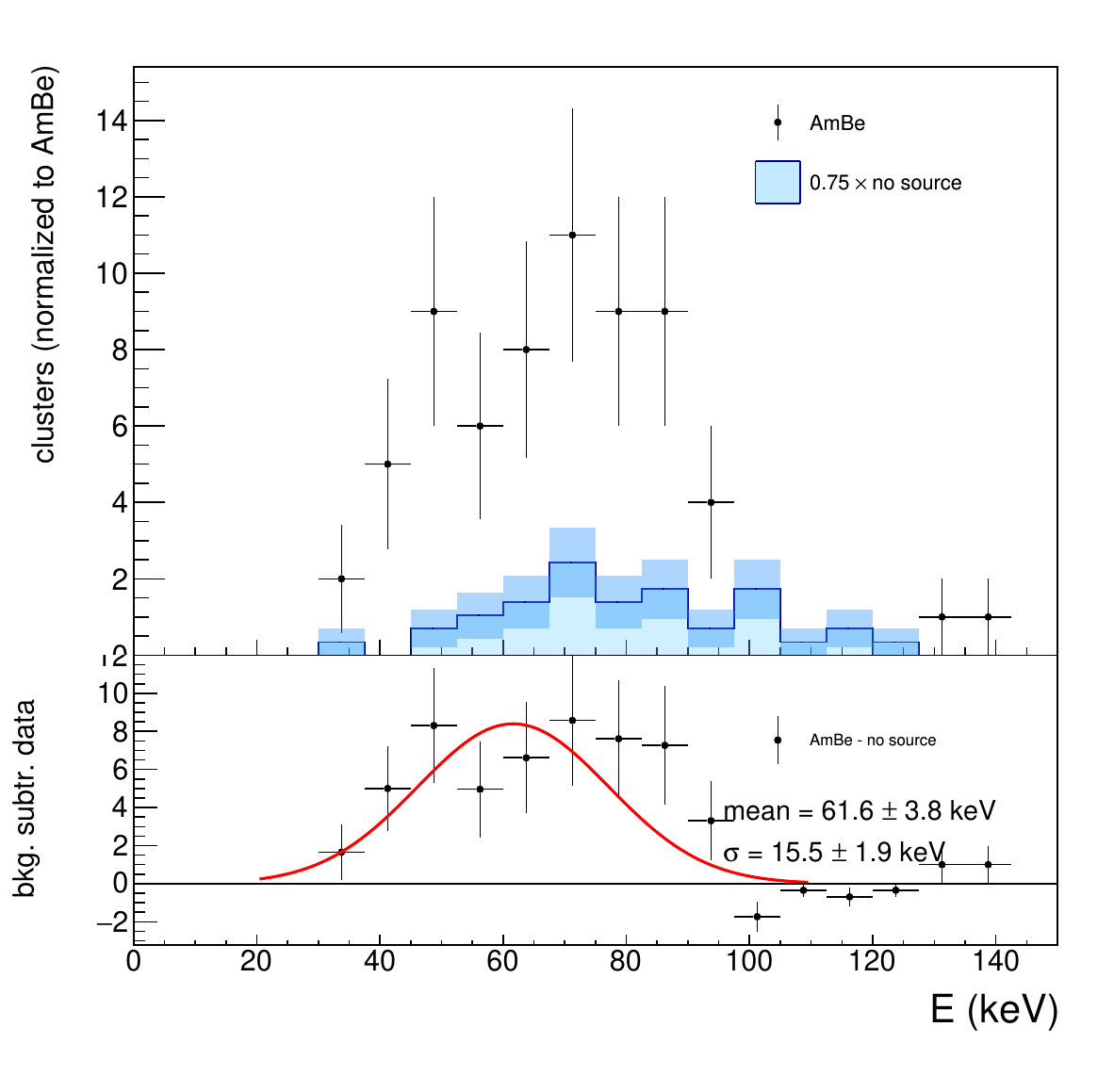}

  \caption{Calibrated energy spectrum for candidates in the control
    region $PR$, defined in the text. The background-subtracted
    distribution is fitted with a Gaussian PDF, which shows a mean
    value compatible with $E=59\keV$ originated from the $^{241}$Am
    $\gamma$s interaction within the gas. The filled dark-azure band
    represents the statistical uncertainty on data with no
    source. \label{fig:59keV}}

  \end{center}
\end{figure}

In addition, it shows a smaller component, also present only in the
data with \ambe source, of clusters with a moderate track length,
$1.5 \lesssim l_p \lesssim 3.0$\unit{cm}, and a lower energy density
than the one characteristic of the nuclear recoils
($9\lesssim\delta\lesssim12$). Since the density is inversely
proportional to the number of active pixels $n_p$, which is correlated
to the track length, the almost linear decrease of $\delta$ as a
function of $l_p$ points to a component with fixed energy. The
$^{241}$Am is expected to produce photons with $E=59\keV$. This
hypothesis is verified by introducing an oblique selection in the
$\delta-l_p$ plane: $\abs{\delta-y}<2$, where $y=14-p_l/50$, for the
clusters with $120<l_p<250$\unit{pixels}, defining the \textit{photon
control region}, $PR$. The approximate oblique region in the
$\delta-l_p$ plane corresponding to $PR$ is also shown in
Fig.~\ref{fig:dvsl}.  The obtained energy spectrum for these clusters
is shown in Fig.~\ref{fig:59keV}, which indeed shows a maximum at
$E=61.6\pm3.8\keV$, within the expected resolution. These events are
thus rejected from the nuclear recoils candidates by vetoing the $PR$
phase space.

\subsection{PMT-based cosmic ray suppression}
\begin{figure}[ht]
  \begin{center}
    \includegraphics[width=0.69\linewidth]{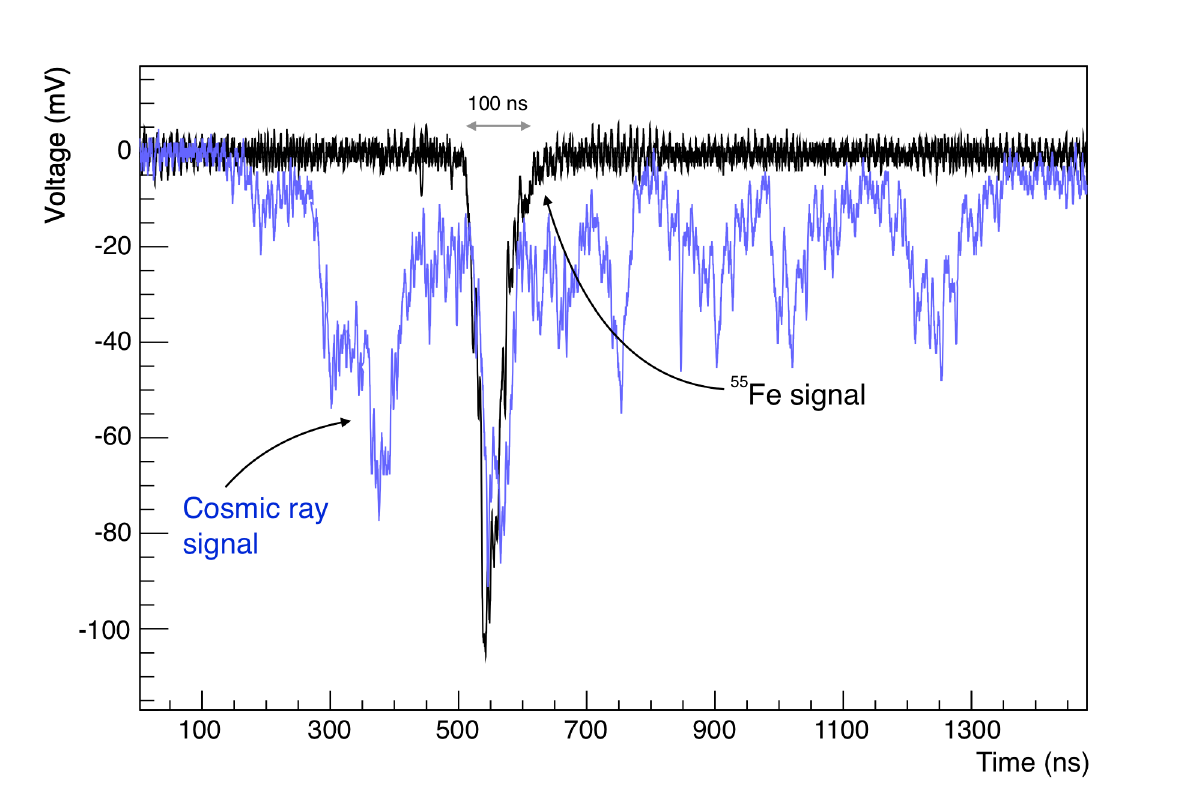}

    \caption{Example of two acquired waveforms: one short pulse
  recorded in presence of \fe radioactive source, together with a long
  signal very likely due to a cosmic ray
  track.  \label{fig:waveforms}}

  \end{center}
\end{figure}

An independent information to the light detected by the sCMOS sensor
of the camera is obtained from the PMT pulse, used to trigger the
image acquisition. For each image acquired, the corresponding PMT
pulse waveform is recorded.  Tracks from cosmic rays, which typically
have a large angle with respect the cathode plane, as shown in
Fig.~\ref{fig:cosmics} (right), show a broad PMT waveform,
characterized by different arrival times of the several ionization
clusters produced along the track at different $z$. Conversely,
spot-like signals like \fe deposits or nuclear recoils are
characterized by a short pulse, as shown in Fig.~\ref{fig:waveforms}.

\begin{figure}[ht]
  \begin{center}
  \includegraphics[width=0.45\linewidth]{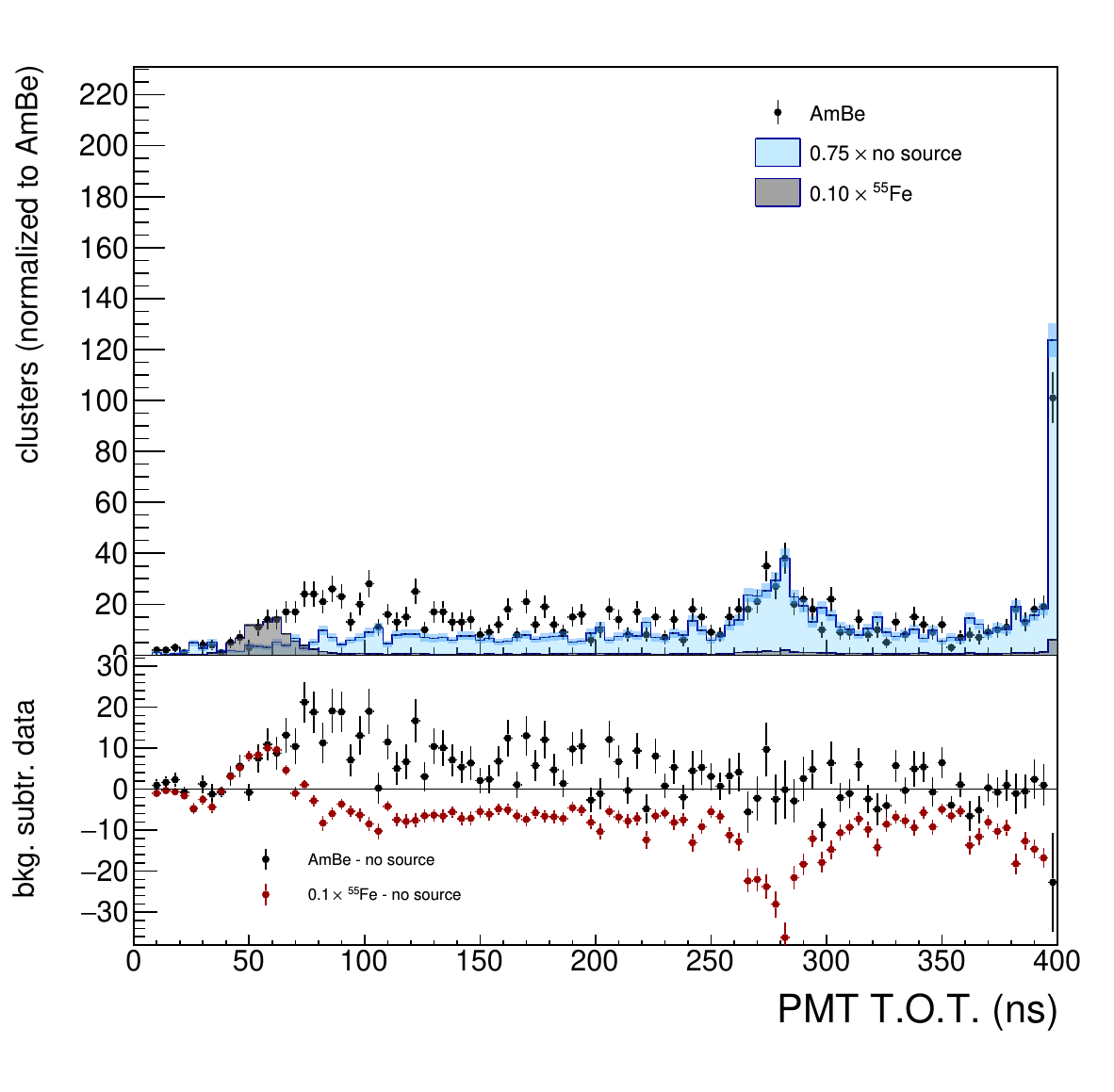}

   \caption{PMT waveform time over threshold ($TOT$).  The last bin
    integrates all the events with $TOT>400$\unit{ns}. Filled points
    represent data with \ambe source, dark gray (light blue)
    distribution represents data with \fe source (no source).  The
    normalization of data without source is to the same exposure time
    of the \ambe one, with trigger scale factor $\varepsilon_{SF}$
    applied. For the data with \fe, a scaling factor of one tenth is
    applied for clearness, given the larger activity of this
    source. Filled dark-grey and dark-azure
    bands represent the statistical uncertainty on data with no source
    and \fe source, respectively \label{fig:pmttot}}

  \end{center}
\end{figure}

\begin{figure}[ht]
  \begin{center}
  \includegraphics[width=0.45\linewidth]{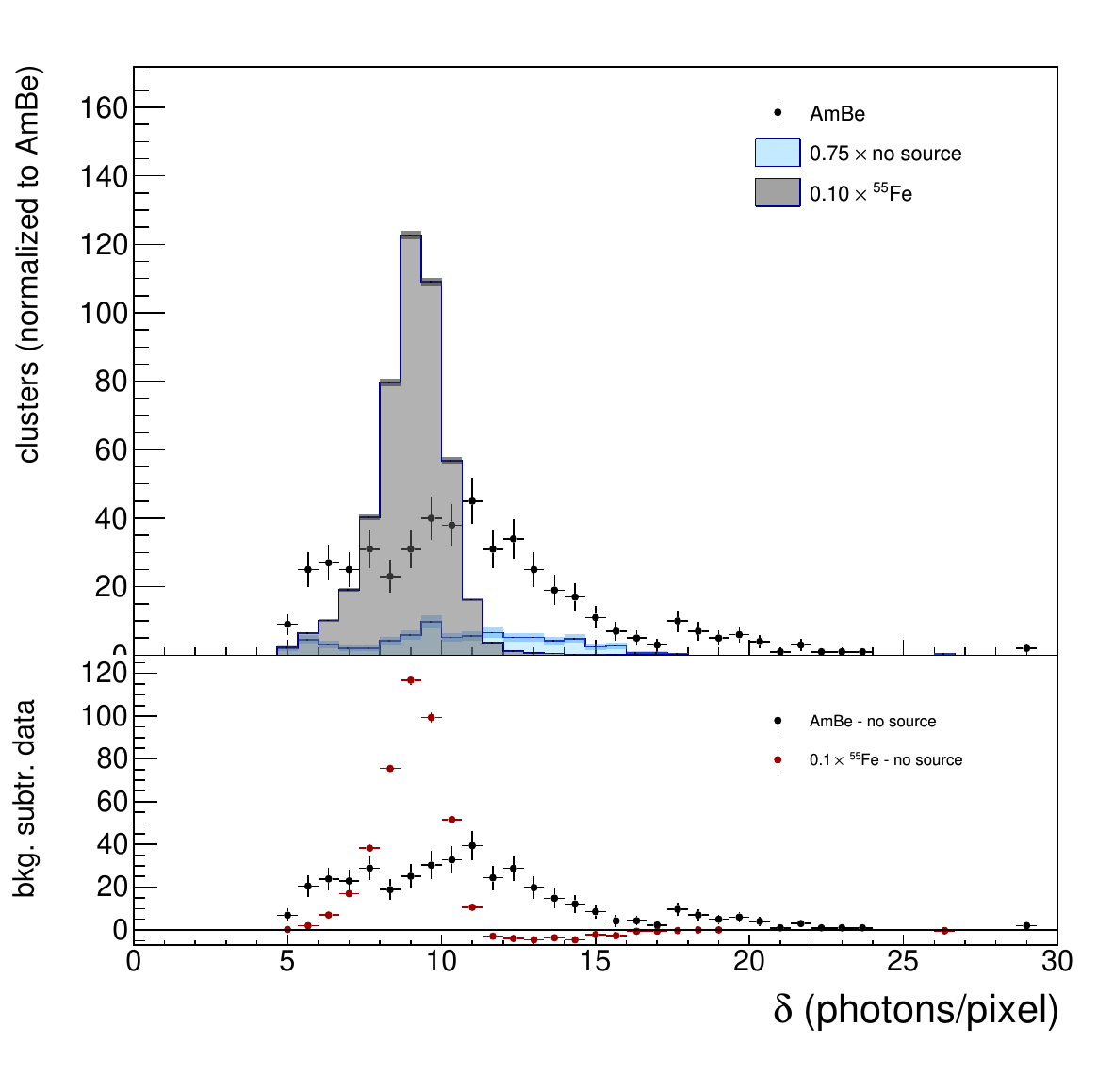}
  \includegraphics[width=0.45\linewidth]{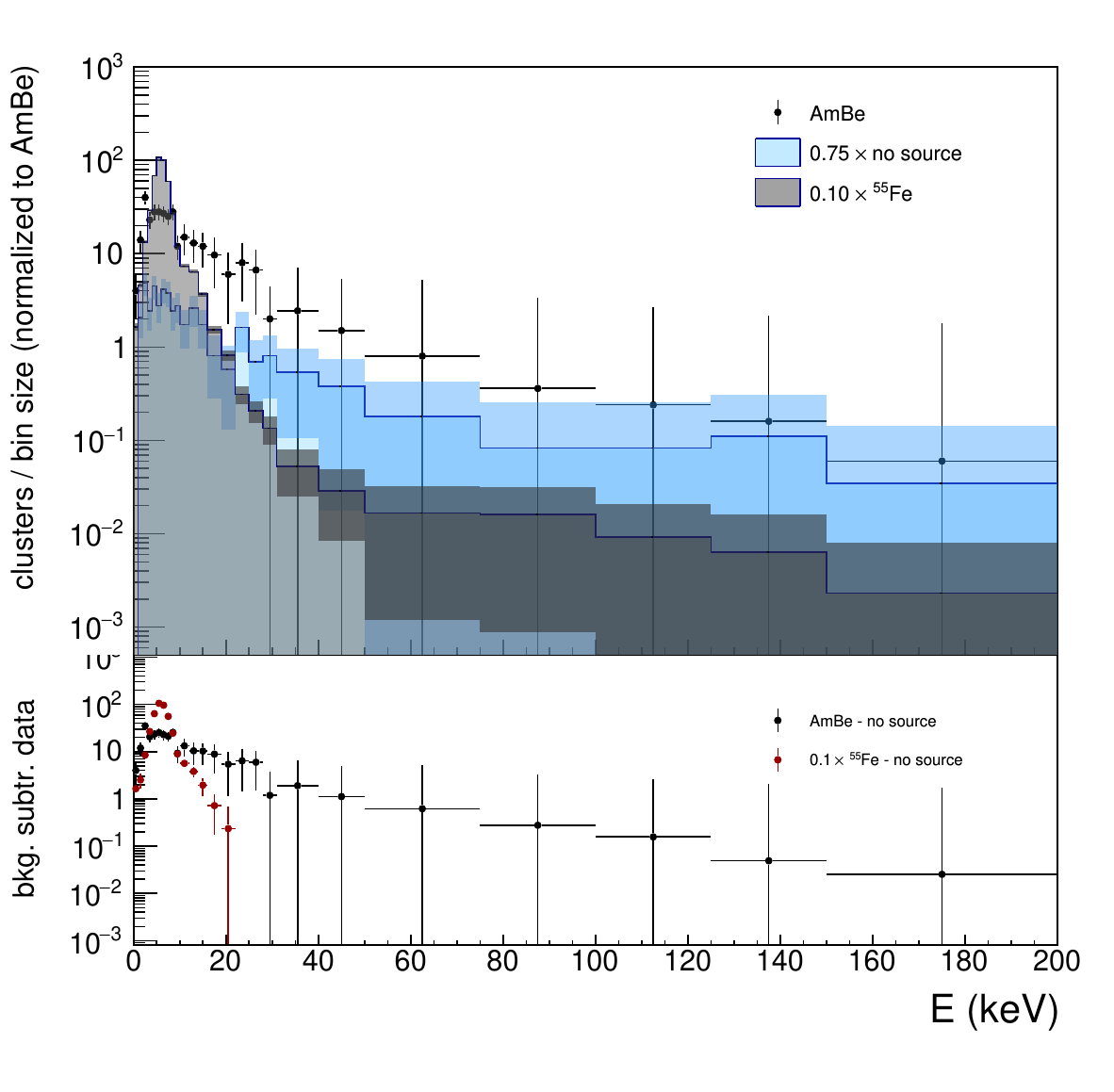}

  \caption{Supercluster light density $\delta$ (left) and calibrated
    energy $E$ (right), after the preselection and cosmic ray
    suppression described in the text to select nuclear recoil
    candidates. Filled points represent data with \ambe source, dark
    gray (light blue) distribution represents data with \fe source
    (no-source).  The normalization of no-source data is to the same
    exposure time of the \ambe data, with the trigger scale factor
    $\varepsilon_{SF}$ applied. For the data with \fe, a scaling
    factor of one tenth is applied for clearness, given the larger
    activity of this source. Filled dark-grey and dark-azure
    bands represent the statistical uncertainty on data with no source
    and \fe source, respectively \label{fig:presel}}

  \end{center}
\end{figure}

The Time Over Threshold (\textit{TOT}) of the PMT pulse was measured,
and is shown in Fig.~\ref{fig:pmttot}. It can be seen from the region
around 270\unit{ns}, dominated by the cosmic rays also in the data
with the \ambe source, that the trigger scale factor
$\varepsilon_{SF}$ also holds for the PMT event rate.
As expected, spot-like clusters (in 3D) correspond to a short pulse in
the PMT, while cosmic ray tracks have a much larger pulse. The
contribution of cosmic ray tracks is clearly visible in the data with
radioactive sources. A selection on this variable is helpful to
further reject residual cosmic rays background present in the \ambe or
\fe data, in particular tracks which may have been split in multiple
superclusters, like the case shown in Fig.~\ref{fig:super_clusters2}
(bottom), and thus passing the above preselection on the cluster
shapes. A selection $TOT<250$\unit{ns} is then imposed.  It has an
efficiency of 98\% on cluster candidates in
\ambe data (after muon-induced background subtraction), while it is only 80\%
efficient on data with \fe source. This larger value is expected
because of the residual contamination of signals from cosmic rays,
which fulfill the selection because their track is split in multiple
sub-clusters, or because they are only partially visible in the sCMOS
sensor image. These can be eventually detected as long, in the time
dimension, by the PMT.  The light density and the energy spectrum of
the preselected clusters are shown in Fig.~\ref{fig:presel}.

\subsection{Light density and  \fe events rejection}
\begin{figure}[ht]
  \begin{center}
  \includegraphics[width=0.45\linewidth]{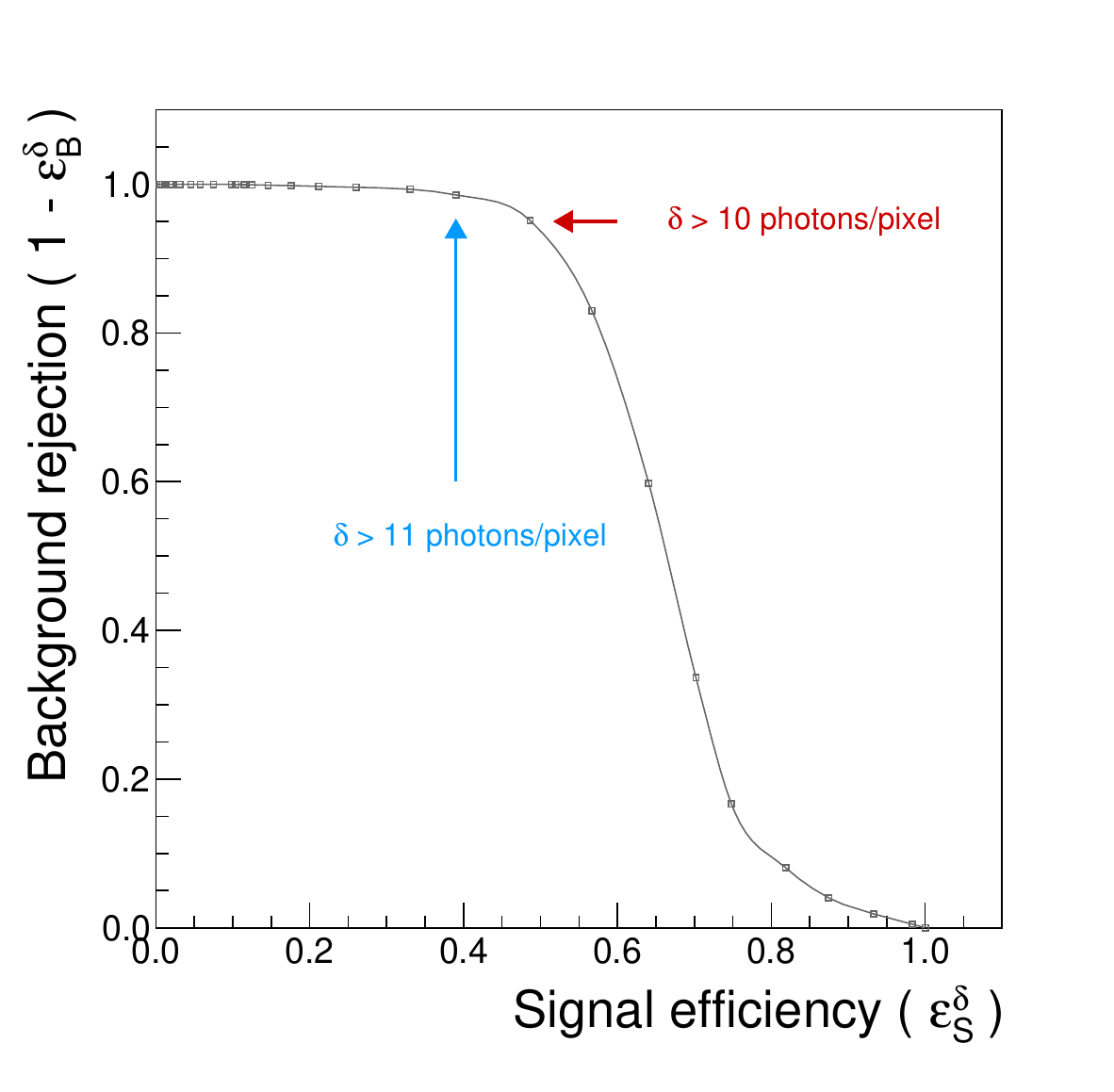}

  \caption{Background rejection as a function of the signal
    efficiency, varying the selection on the $\delta$ variable in data
    with either \fe (background sample) or \ambe (signal sample)
    sources.  \label{fig:roc}}

  \end{center}
\end{figure}

The light density distributions, after the above preselection and
cosmic ray suppression, for the data taken with \ambe source, data
with \fe source, and data without any artificial source, are
different.  The cosmic-background-subtracted distributions of $\delta$
in \ambe data and \fe data, shown in the bottom panel of
Fig.~\ref{fig:presel} (left), are used to evaluate a curve of 5.9\keV
electron recoils rejection ($1-\varepsilon^\delta_{B}$) as a function
of signal efficiency ($\varepsilon^\delta_{S}$), obtained varying the
selection on $\delta$, shown in Fig.~\ref{fig:roc}.  The same
procedure could be applied to estimate the rejection factor against
the cosmic-ray-induced background, but this is not shown because of
the limited size of the no-source data. This kind of background will
however be negligible when operating the detector underground, in the
context of the \cygno project, so no further estimates are given for
this source.

Table~\ref{tab:roc} shows the full signal efficiency and electrons
rejection factor for two example working points, $\mathrm{WP}_{40}$
and $\mathrm{WP}_{50}$, having 40\% and 50\% signal efficiency for the
selection on $\delta$, averaged over the full energy spectrum
exploited in the \ambe data. They correspond to a selection
$\delta>11$ and $\delta>10$, respectively.  While this cut-based
approach is minimalist, and could be improved by profiting of the
correlations among $\delta$ and the observables used in the
preselection in a more sophisticated multivariate analysis, it shows
that a rejection factor approximately in the range
[$10^{-3}$-$10^{-2}$] of electron recoils at $E=5.9\keV$ with a
gaseous detector at atmospheric pressure can be obtained, while
retaining a high fraction of signal events.

\begin{table*}[t]

\caption{Signal (nuclear-recoil-induced by \ambe radioactive source) and background (photo-electron recoils of X-rays
         with $E=5.9$\keV from \fe radioactive source) efficiency for
         two different selections on $\delta$.\label{tab:roc}}

\vspace{10pt}
\normalsize
\centering
\begin{tabular}{l c c c | c c c }
  \hline\hline
  working point & \multicolumn{3}{c}{Signal efficiency} & \multicolumn{3}{c}{Background efficiency} \\
  \hline
  & $\varepsilon_{S}^{presel}$ & $\varepsilon_{S}^{\delta}$ & $\varepsilon_{S}^{total}$ & $\varepsilon_{B}^{presel}$ & $\varepsilon_{B}^{\delta}$ & $\varepsilon_{B}^{total}$ \\
  \hline
  $\mathrm{WP}_{50}$  & 0.98                        & 0.51                      & 0.50                     & 0.70                     & 0.050                     & 0.035 \\
  $\mathrm{WP}_{40}$  & 0.98                        & 0.41                      & 0.40                     & 0.70                     & 0.012                     & 0.008 \\
  \hline\hline
\end{tabular}
\end{table*}

\subsection{Nuclear recoil energy spectrum and differential efficiency}

\begin{figure}[ht]
  \begin{center}
    \includegraphics[width=0.45\linewidth]{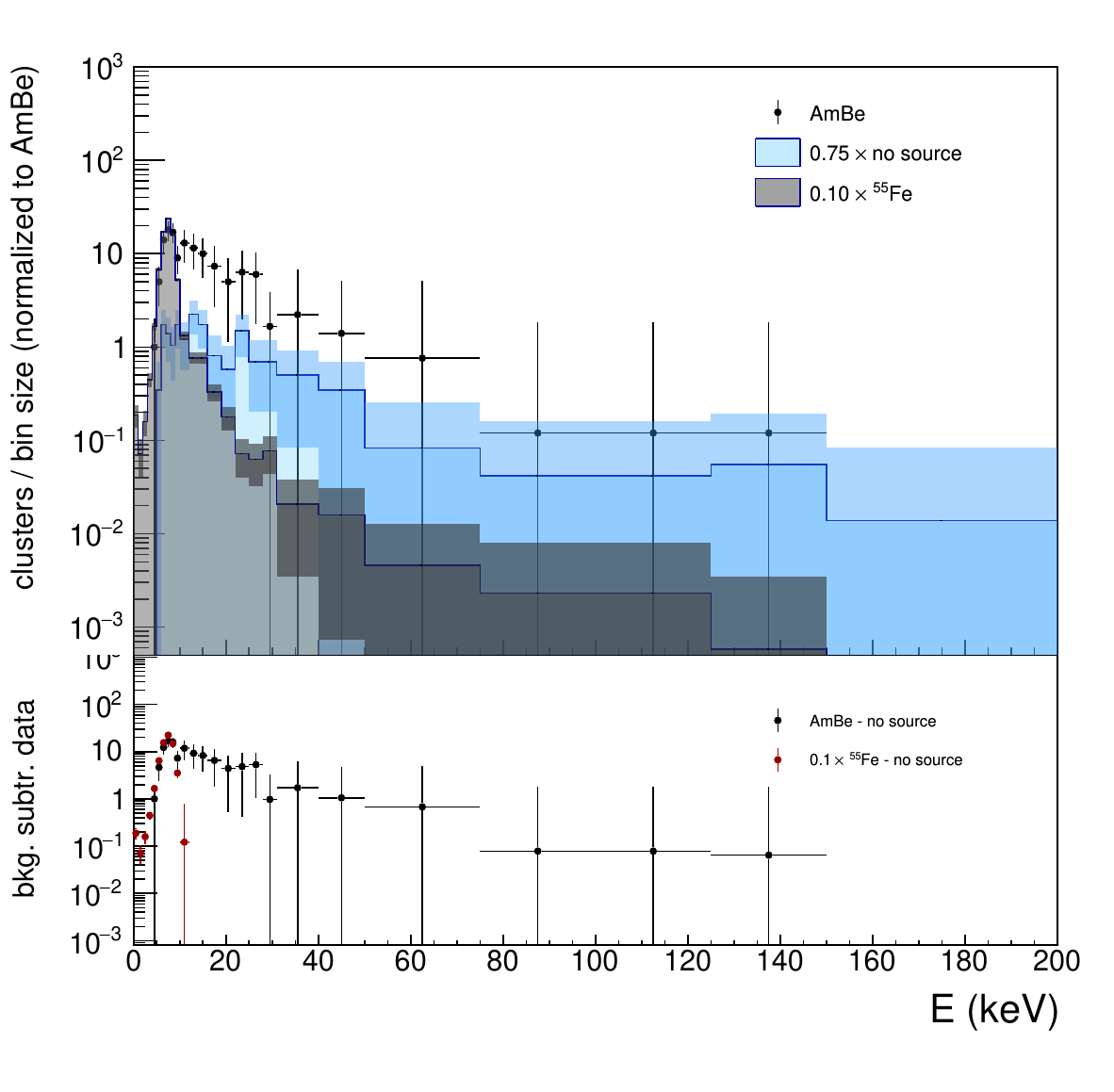}
    \includegraphics[width=0.45\linewidth]{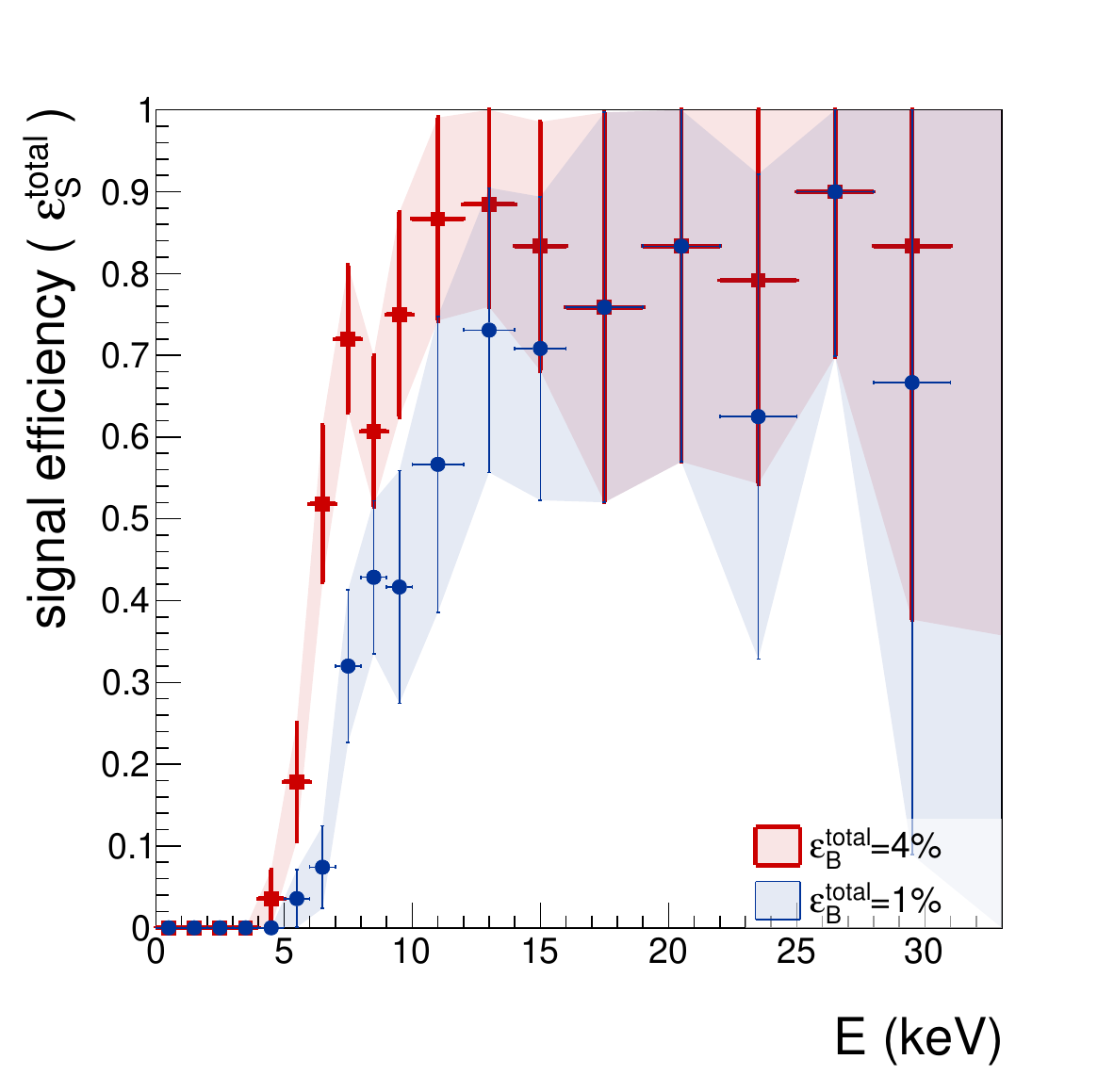}

    \caption{Left: supercluster calibrated energy $E$ (left), after
      the full selection, which includes $\delta>10$, 50\% efficient
      on signal, to select nuclear recoil candidates. Filled points
      represent data with \ambe source, dark gray (light blue)
      distribution represents \fe source (no-source) data.  The
      normalization of no-source data is to the same exposure time of
      the \ambe data, with the trigger scale factor $\varepsilon_{SF}$
      applied. For the \fe data, a scaling factor of one tenth is
      applied for clearness, given the larger activity of this
      source. Filled dark-grey and dark-azure bands represent the
      statistical uncertainty on data with no source and \fe source,
      respectively. Right: efficiency for nuclear recoil candidates as
      a function of energy, estimated on \ambe data, for two example
      selections, described in the text, having either 4\% or 1\%
      efficiency on electron recoils at
      $E=5.9\keV$. \label{fig:fullsel_effi}}

  \end{center}
\end{figure}

The energy spectrum for the candidates with
$\varepsilon_{S}^{total}$=50\% in the \ambe sample is shown in
Fig.~\ref{fig:fullsel_effi} (left).  The signal efficiency is then
computed for both the example working points in bins of the visible
energy. The efficiency, $\varepsilon_{B}^{total}$, represents a
$\gamma$ background efficiency at a fixed energy $E=5.9\keV$, \ie, the
energy of the photons emitted by the \fe source. For the
$\mathrm{WP}_{50}$, the efficiency for very low-energy recoils,
$E=5.9\keV$, is still 18\%, dropping to almost zero at
$E\lesssim4\keV$. The efficiencies reported here only account for
analysis selection efficiency, while an absolute trigger efficiency
should be also considered.  Since the trigger will be probably
different in the final CYGNO experiment, where the cosmic-ray-induced
background would be much reduced, this estimate is not attempted for
this study.

\begin{figure}[ht]
  \begin{center}
  \includegraphics[width=0.49\linewidth]{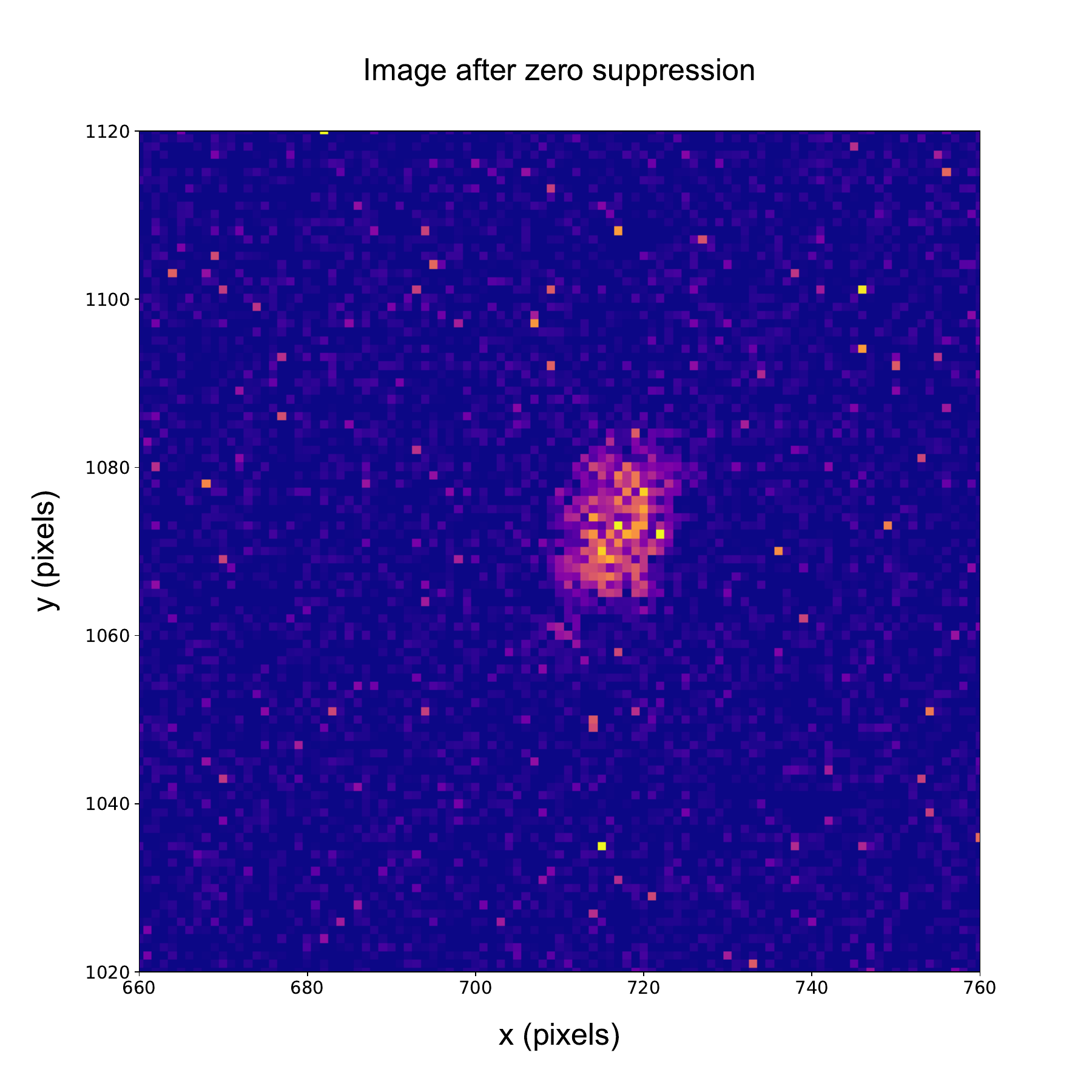}
  \includegraphics[width=0.49\linewidth]{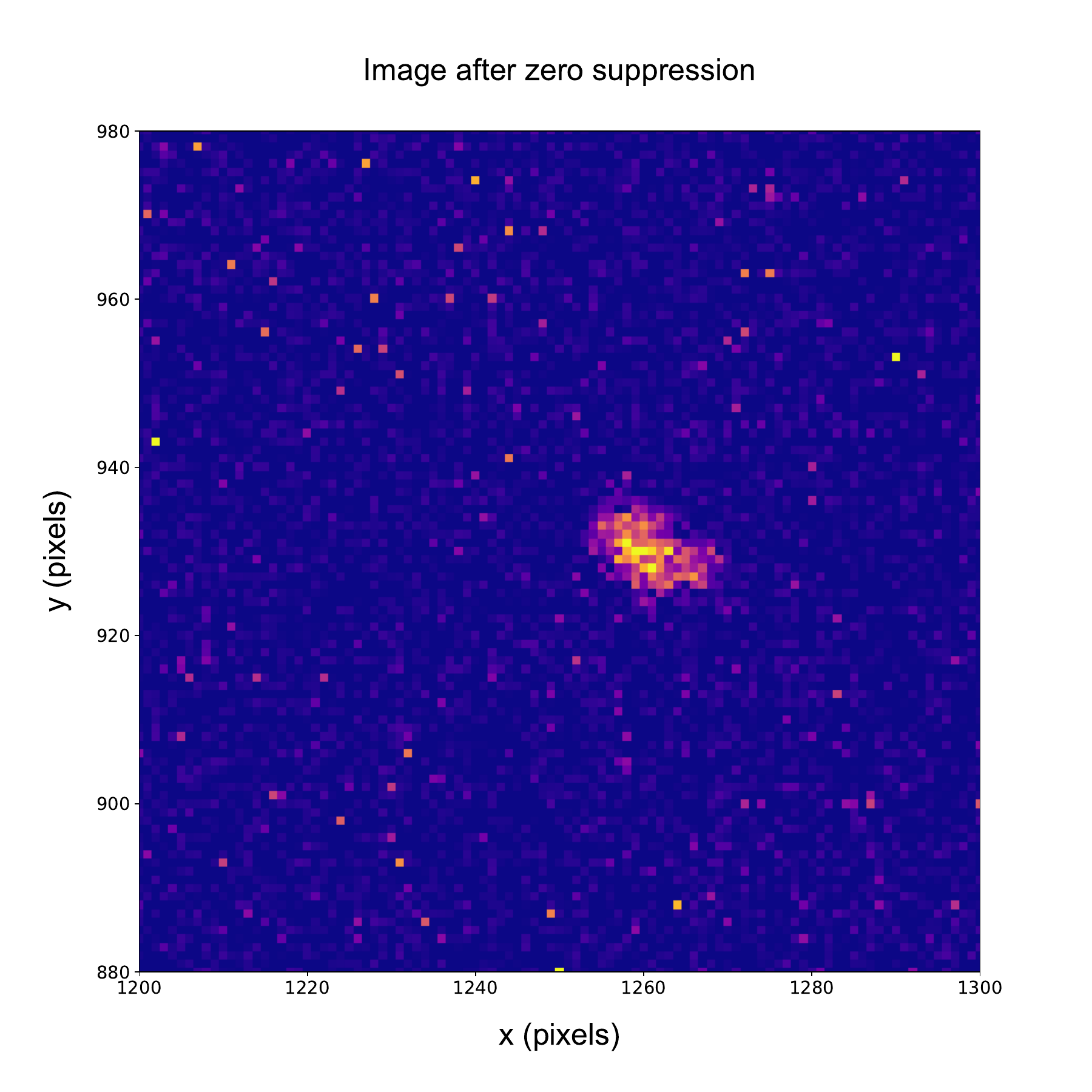}

  \caption{Examples of two nuclear recoil candidates, selected with  the full selection, shown in a portion of $100\times100$ pixel matrix, after the zero suppression of the image. Left: a candidate
    with $E=5.2\keV$ and $\delta=10.5$, right: a candidate with
    $E=6.0\keV$ and $\delta=10$.  \label{fig:lowEnergyNR}}

\end{center}
\end{figure}

Two candidate nuclear recoils images, fulfilling the
$\mathrm{WP}_{50}$ selection (with a light density $\delta\gtrsim10$
photons/pixels and with energies of 5.2 and 6.0\keV) are shown in
Fig.~\ref{fig:lowEnergyNR}. The displayed images are a portion of the
full-resolution frame, after the pedestal subtraction. While the
determination of the direction of detected nuclear recoil is still
under study, it appears pretty clear from the image that some
sensitivity to their direction, even at such low energies, is retained
and can be further exploited.

\clearpage

 \section{Conclusion and outlook}

A method to efficiently identify recoiling nuclei after an elastic
scattering with fast neutrons with an optically readout TPC was
presented in this paper.  A 7 liter prototype was employed by exposing
its sensitive volume to two kinds of neutral particles in an
overground location:
\begin{itemize}
\item photons with energy of 5.9~\keV and 59~\keV respectively
  provided by a radioactive source of \fe and by one of $^{241}$Am
  able to produce electron recoils with equal energy by means of
  photoelectric effect;
\item neutrons with kinetic energy of few MeV produced by an \ambe
  source that can create nuclear recoils with kinetic energy lower
  than the neutron ones.
\end{itemize}

The high sensitivity of the adopted sCMOS optical sensor allowed a
very good efficiency in detecting events with an energy released in
gas even below 10~\keV.

Moreover, the possibility of exploiting the topological information
(shape, size and more) of clusters of emitted light allowed to develop
algorithms able to reconstruct not only the total deposited energy,
but also to identify the kind of the recoiling ionizing particles in
the gas (either an electron or a nucleus). Cosmic ray long tracks are
also clearly separated.

Because of their larger mass and electric charge, nuclear recoils are
expected to release their energy by ionizing the gas molecules in a few
hundred \unit{$\mu$m} while the electrons are able to travel longer
paths. For this reason, by exploiting the spatial distribution of the
collected light, it was possible to identify 5.9\keV electron recoils
with an efficiency of 96.5\% (99.2\%) against nuclear recoils by
retaining a capability of detecting them with an efficiency of 50\%
(40\%), averaged across the measured \ambe spectrum.

In particular, the nuclear recoil detection efficiency was measured to
be 40\% for deposited energies lower than 20\keV and 14\% in the range
(5--10)\keV.

The results obtained in the studies presented in this paper can be
improved by means of more sophisticated analyses exploiting a
multivariate approach, which combines a more complete topological
information about the light distribution along the tracks.  Additional
enhancement of sensitivity can be achieved with a DAQ system
collecting single PMT waveforms to be correlated with the track
reconstructed in the sCMOS images.

\section{Acknowledgments}
We are grateful to ``Servizio Sorgente LNF''.  This work was supported
by the European Research Council (ERC) under the European Union’s
Horizon 2020 research and innovation program (grant agreement No
818744).

\bibliographystyle{ieeetr}
\bibliography{LEMON-20-001}

\end{document}